\documentstyle[epsfig,seceq,twocolumn]{jpsj}
\title
{
Study of Magnetic Excitation in Singlet-Ground-State Magnets CsFeCl$_3$ and RbFeCl$_3$ by Nuclear Magnetic Relaxation
}
\def\runtitle
{
NMR Study on CsFeCl$_3$ and RbFeCl$_3$       
}
\author
{
Mitsuru {\sc Toda}\footnote{E-mail: mtoda@rri.kyoto-u.ac.jp}, Takao {\sc Goto$^1$}, Meiro {\sc Chiba$^2$} and Naoshi {\sc Suzuki$^3$}}
 
\def\runauthor
{
 M. Toda, et al.             
}

\inst
{Research Reactor Institute, Kyoto University, Kumatori 590-0494 \\
$^1$ Graduate School of Human and Environmental Studies, Kyoto University, Kyoto 606-8501 \\
$^2$ Department of Applied Physics, Fukui University, Fukui 910-8507 \\
$^3$ Department of Physical Science, Graduate School of Engineering Science, Osaka University, Toyonaka 560-8531
}

\abst
{
The temperature dependences of spin-lattice relaxation time $T_1$ of $^{133}$Cs in CsFeCl$_3$ and $^{87}$Rb in RbFeCl$_3$ were measured in the temperature range between 1.5 K and 22 K, at various fields up to 7 T applied parallel (or perpendicular) to the $c$-axis, and the analysis was made on the basis of the DCEFA. The mechanism of the nuclear magnetic relaxation is interpreted in terms of the magnetic fluctuations which are characterized by the singlet ground state system. In the field region where the phase transition occurs, $T_1^{-1}$ exhibited the tendency of divergence near $T_{\rm N}$, and this feature was ascribed to the transverse spin fluctuation associated with the mode softening at the $K$-point. It was found that the damping constant of the soft mode is remarkably affected by the occurrence of the magnetic ordering at lower temperature, and increases largely in the field region where the phase transition occurs.
} 
\kword{singlet ground state, softening effect, nuclear spin-lattice relaxation, damping constant,CsFeCl$_3$,RbFeCl$_3$}

\begin{document}
\sloppy
\maketitle
\section{Introduction}
 Hexagonal compounds  CsFeCl$_3$ and RbFeCl$_3$ are typical singlet ground state magnets. The magnetic system of Fe$^{2+}$ ion which is described in terms of a fictitious spin of $S = 1$, is characterized by the singlet ground state and the excited doublet states due to a large positive single-ion anisotropy along the crystallographic $c$-axis. Hereafter, these two compounds are abbreviated as CFC and RFC, respectively. Because of the competition between this anisotropy and the dominant ferromagnetic intra-chain exchange interaction along the $c$-axis, these compounds exhibit various magnetic properties. The common features are the softening of magnetic excitations, which gives rise to 3-dimensional long-range order (3D-LRO) such as the commensurate and incommensurate phases. On the other hand, there appears different features originating from a difference in the relative magnitude of the anisotropy with respect to the intra-chain exchange interaction, even if it is small.

 So far, magnetic properties such as an excitation spectrum including the characteristic softening effect and a phase transition, have been studied extensively by the measurements of magnetization\cite{rf:35,rf:31,rf:7,rf:32,rf:23,rf:34}, specific heat\cite{rf:1,rf:23} and inelastic neutron scattering\cite{rf:8,rf:6,rf:3}. The phase diagrams have been determined by the specific heat measurements for both $H {\mbox {\scriptsize $/\!\!/$}} c$\cite{rf:1} and $H$$\perp$$c$\cite{rf:23}. In the case of CFC, the system remains non-magnetic down to zero K at zero field. While, when the external field is applied along the $c$-axis, one of the excited  doublet decreases with increasing field and the energy levels of the ground state and the excited state cross each other at the field $H_c$ (=7.5 T).  Then the field induced ordered state appears in the field range of 4 T$\le$$H$$\le$11 T around $H_c$\cite{rf:7,rf:34,rf:1,rf:8,rf:6,rf:3}. In the case of RFC, the magnetic phase transition occurs at zero field at $T_{\rm N}\sim $2.6 K, and the ordered state disappears when the value of the external field is beyond 2.2 T in case of $H$$\perp$$c$\cite{rf:35,rf:31,rf:32,rf:23,rf:8}. 

 In particular, it should be noted that the softening of magnetic excitation was observed in the inelastic neutron scattering experiments in RFC\cite{rf:8}. It was recognized that the phase transition is driven by a specific mode of magnetic excitation which decreases to zero at the phase transition point ($T_{\rm N}$) with the ordering vector ${\mib q}=(1/3,1/3,0)$ (${\it K}$-point). Suzuki developed the dynamical correlated-effective-field approximation (DCEFA)\cite{rf:20} focusing on the problem of softening, on the basis of the molecular field approximation, by taking into account the effect of spin-correlation and fluctuation. However, it is difficult to obtain the information of the excitation spectrum and the damping of the excitation under the high external field in the neutron scattering experiments. It is interesting to investigate the effect of softening of magnetic excitation and the spin dynamics in the ordering process in the singlet ground state system. 

 In these compounds, the magnetic nuclear spin-lattice relaxation time $T_1$ is expected to be a useful probe for the study of spin dynamics of magnetic systems associated  with low lying excitations caused by the softening. Previously, the temperature and field dependences of $T_1$ of $^{87}$Rb have been measured in RFC\cite{rf:27}. The effect of the soft mode was found in the temperature dependence of the relaxation rate $T_1^{-1}$. Although the behavior of $T_1^{-1}$ was explained qualitatively, quantitative understanding was not necessarily satisfactory. As for CFC, the field dependence of $T_1$ of $^{133}$Cs have been measured up to 14 T ($H {\mbox {\scriptsize $/\!\!/$}} c$) at the constant temperature of 4.2 K in the paramagnetic region\cite{rf:7}. The remarkable field dependence of $T_1^{-1}$ was found with the maximum value at $H_c$ in the field region where the magnetic phase transition occurs. The experimental data were interpreted from the viewpoint of the direct process, which occurs via the overlapping of the ground state and excited state due to the intra-chain exchange interaction. However, the presence of relatively weak temperature dependence in $T_1^{-1}$ has remained unsolved.
 
 The purpose of the present work is twofold: It is to clarify the relaxation mechanism associated with the softening, and then to obtain the information of the softening of magnetic excitation in the paramagnetic state of the singlet ground state system. We have measured the temperature dependence of $T_1^{-1}$ of $^{133}$Cs in CFC, in the temperature range from 1.5 K to 22 K mainly applied parallel to the $c$-axis up to 7 T. Preliminary results were already published\cite{rf:28}. We have also measured the temperature dependence of $T_1^{-1}$ of $^{87}$Rb in RFC in the temperature range from 1.5 K to 22 K mainly applied perpendicular to the $c$-axis up to 3.3 T. The analysis is performed on the basis of the theory of DCEFA developed by Suzuki. We shall survey the experimental procedures in $\S$2 and results in $\S$3. The derivation of the theoretical equation of $T_1^{-1}$ is explained in $\S$4, and $\S$5 is devoted on the discussions over the obtained results. Finally, the conclusion is summarized in $\S$6.
 
\section{Experimental procedures}
 Single crystals of CFC(RFC) were made by mixing the equimolar materials of CsCl(RbCl) and FeCl$_2$. The source materials  were qualified for the purity of 99.9 $\%$. Because of the hygroscopic nature of FeCl$_2$, the materials were dehydrated using HCl gas in the ampoule\cite{rf: 4}. After the dehydration procedure, single crystals were grown by the vertical-gradient-freeze Bridgman method. The sealed ampoule was lowered by winding down a siltex code at a speed of 2cm/day. Single crystals as large as $4\sim5$ ${\rm cm}^3$ were obtained for the maximum size.
 
 The measurements of NMR were performed using the conventional phase-coherent pulsed-NMR spectrometer, which covers frequency range of $5\sim300$ MHz. The NMR spectra were obtained by observing a spin echo signal as a function of an external field. The spin lattice relaxation time $T_1$ was measured by observing a recovery of the signal intensity, after the saturation of nuclear magnetization by comb pulses.

 We determined the nuclear spin lattice relaxation time $T_1$ of $^{87}$Rb and $^{133}$Cs as follows. As for $^{87}$Rb in RFC, the recovery curve of the nuclear magnetization ($I=3/2$) is given as follows, when the quadrupole splitting is large\cite{rf:9}; 
\begin{equation}
\frac{M_0-M(t)}{M_0}= a\cdot\exp(-\frac{t}{T_1})+b\cdot\exp(-\frac{6t}{T_1}).
\label{twocomp}
\end{equation}
Here the values of coefficients $a$ and $b$ were chosen as adjustable parameters, because it was difficult to obtain the complete saturation of magnetization. Reflecting the fact that the quadrupole splitting is sufficiently large compared to the spectral line width, the spectrum splits into three lines. 
\begin{figure}[t]
  \begin{center}
    \epsfxsize=8.5cm    %  To keep the "height/width" ratio of your own figure,
    \epsfysize=6.0cm     %  please use either "epsfxsize" or "epsfysize" command.
    \epsfbox{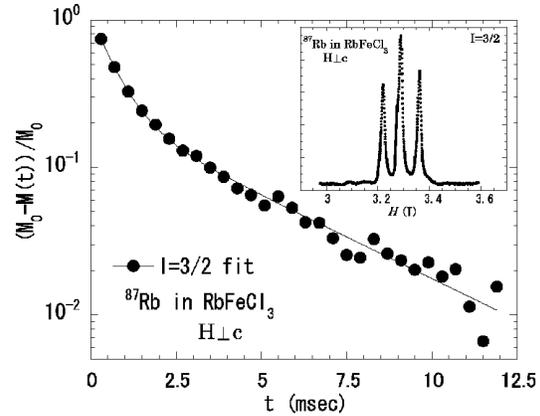}
  \end{center}
\caption{A recovery of the intensity of a spin echo signal of $^{87}$Rb in the paramagnetic state at $T$=2.8 K and $H$=3.33 T ($H \bot c$). The solid line is the fit to eq.(\ref{twocomp}) in the text. The inset shows the whole NMR spectrum of $^{87}$Rb.}
\label{6}
\end{figure}
\begin{figure}[b]
  \begin{center}
    \epsfxsize=8.5cm    %  To keep the "height/width" ratio of your own figure,
    \epsfysize=6.2cm     %  please use either "epsfxsize" or "epsfysize" command.
    \epsfbox{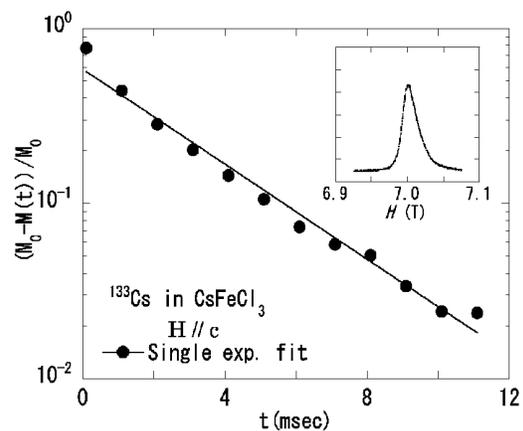}
  \end{center}
  \caption{A recovery of the intensity of a spin echo signal of $^{133}$Cs in the paramagnetic state at $T$=20 K and $H$=7 T ($H {\mbox {\scriptsize $/\!\!/$}} c$). The inset shows the NMR spectrum of $^{133}$Cs.}%{}内にタイトルを記入してください
  \label{1}
\end{figure}
Figure \ref{6} shows a recovery curve of the intensity of a spin echo signal measured at the central line, and the whole NMR spectrum is shown in the inset. As is seen, the obtained curve is well fitted to eq.(\ref{twocomp}).
 On the other hand, in the case of CFC, the recovery curve is almost exponential as shown in Fig.\ref{1}. The inset in Fig.\ref{1} represents the NMR spectrum of $^{133}$Cs. Since the quadrupole moment of $^{133}$Cs is extremely small, one absorption line was observed as shown in the inset of Fig.\ref{1}, and the value of $T_1$ is determined by fitting to the single-exponential recovery equation. 

\section{Experimental results}
 
\begin{figure}[h]
 \vspace{10pt}
  \begin{center}
    \epsfxsize=8.5cm    %  To keep the "height/width" ratio of your own figure,
    \epsfysize=9.8cm     %  please use either "epsfxsize" or "epsfysize" command.
    \epsfbox{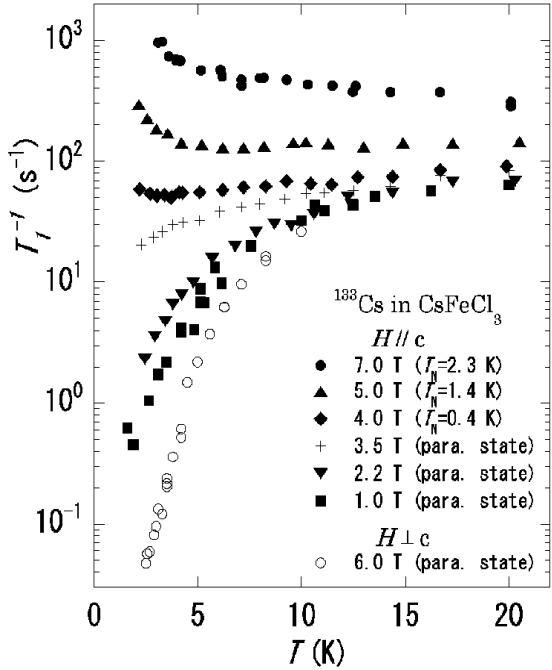}
  \end{center}\caption{The temperature dependences of $T_1^{-1}$ of $^{133}$Cs at various fields applied parallel (or perpendicular) to the $c$-axis.}
  \label{3}
\end{figure}

Figure \ref{3} shows the temperature dependences of the relaxation rate $T_1^{-1}$ for CFC at various fields up to 7 T ($H {\mbox {\scriptsize $/\!\!/$}} c$), and at 6.0 T ($H \bot c$).
 As is seen, there is a distinct difference in the temperature dependences of $T_1^{-1}$ between the lower ($H$$\le$3 T) and higher ($H$$\ge$3 T) field regions. At low fields, $T_1^{-1}$ decreases remarkably with decreasing temperature. As the field increases, the temperature dependence of $T_1^{-1}$ becomes more moderate, and this suggests that the gap energy decreases. As is shown in the open circle in Fig. \ref{3},  the value of $T_1^{-1}$ for $H \bot c$ exhibits remarkable decrease with decreasing temperature reflecting the fact that the gap does not close. On the other hand, in the higher field region, $T_1^{-1}$ is almost temperature-independent down to 5 K, and exhibits an appreciable increase at lower temperatures. It seems that such a qualitative difference corresponds just to the absence and the realization of a magnetic phase transition. 
\begin{figure}[b]
 \vspace{10pt}
  \begin{center}
    \epsfxsize=8.5cm    %  To keep the "height/width" ratio of your own figure,
    \epsfysize=8.0cm     %  please use either "epsfxsize" or "epsfysize" command.
    \epsfbox{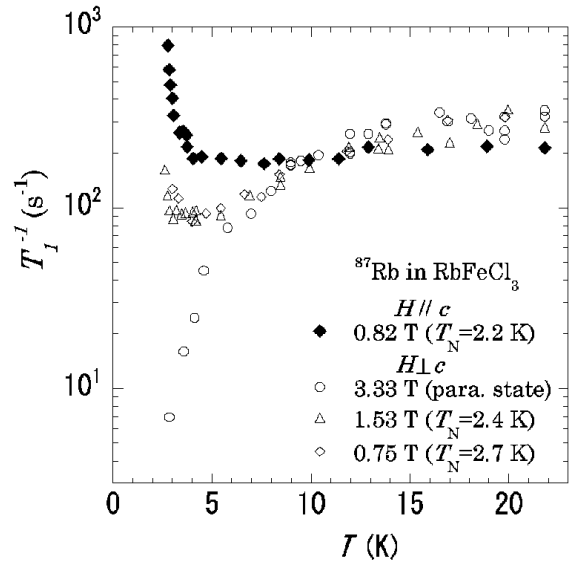}
  \end{center}
\caption{The temperature dependences of $T_1^{-1}$ of $^{87}$Rb at various fields applied parallel (or perpendicular) to the $c$-axis.}
\label{8}
\end{figure} 

Figure \ref{8} shows the temperature dependences of $T_1^{-1}$ of $^{87}$Rb in RFC. In this compound, a phase transition occurs under zero field at 2.6 K. As shown by the data obtained at 0.82 T ($H {\mbox {\scriptsize $/\!\!/$}} c$), $T_1^{-1}$ exhibits a drastic increase around the transition temperature. Above about 5 K, the values of $T_1^{-1}$ are almost constant with the value of $T_1^{-1} \approx 200$ ${\rm s}^{-1}$. In the case of $H \bot c$, the transition temperature decreases to zero K at 2.2 T with increasing external field. The relaxation rate measured at 0.75 T and 1.53 T shows drastic increase around the phase transition temperature. It was found that $T_1^{-1}$ increases gradually with increasing temperature above 5 K. While, as shown in the data taken at 3.33 T,  $T_1^{-1}$ for the field above the critical field shows drastic decrease with decreasing temperature.

 As a common feature in both compounds, the relaxation rate decreases remarkably with decreasing temperature in the field range where the phase transition does not occur. On the other hand, in the field region where the phase transition occurs, the relaxation rate exhibits the divergent behavior. In the following, we shall analyze the above experimental results on the basis of the theory of the DCEFA.

\section{Theoretical background}
\subsection{Description of electron spin system}
In the DCEFA theory, the spin product is decoupled in the following manner,
\begin{eqnarray}
\hspace{-3mm} & &S_{i\mu}S_{j\mu}\to S_{i\mu} (\left\langle S_{j\mu} \right\rangle+\alpha (S_{i\mu}-\left\langle S_{i\mu} \right\rangle)) \nonumber \\
&+& S_{j\mu} (\left\langle S_{i\mu} \right\rangle+\alpha (S_{j\mu}-\left\langle S_{j\mu} \right\rangle)),\hspace{3mm} (\mu=x,y,z)
\end{eqnarray}
 where $\left\langle S_{i\mu} \right\rangle$ and $\left\langle S_{j\mu} \right\rangle$ are the thermal averages, and $\alpha$ represents a correlation parameter to be determined self-consistently. Due to the crystalline field, the electron spin system is largely influenced by the direction of $H$ with respect to the anisotropy axis, the $c$-axis. In view of this, we treat individually the electron spin systems in the two cases of $H {\mbox {\scriptsize $/\!\!/$}} c$ and $H \bot c$. In the followings, we define the $x$-axis and $z$-axis as the crystalline $a$-axis and $c$-axis, respectively, in the hexagonal structure of CFC and RFC.
 
\subsection{$H {\mbox{\footnotesize $/\!\!/$}} c$}
First we consider the case of $H {\mbox{\footnotesize $/\!\!/$}} c$. The effective single-ion Hamiltonian of the $i$-th site is expressed as \cite{rf:22}
\begin{equation}
H_{i}^{\it eff}=ES_{iz}^2 +B^{{\mbox{\tiny $/\!\!/$}}} S_{iz},
\label{effsingle}
\end{equation}
with $E=D+\alpha(J^{\perp}(0)-J^{\mbox{\tiny $/\!\!/$}}(0))$ and $B^{\mbox{\tiny $/\!\!/$}}=g_{\mbox{\tiny $/\!\!/$}}\mu_BH + 2J^{\mbox{\tiny $/\!\!/$}}(0)(1-\alpha)\langle S_z \rangle$. Here, $\langle S_z \rangle $ is the thermal average of the induced moment which is independent of the $i$-th site; $J^{\gamma}(0)=J_1^{\gamma}+3J_2^{\gamma}$ ($\gamma=\bot$, ${\mbox{\footnotesize $/\!\!/$}}$) represents ${\mib q}=0$ component of the Fourier-transform $J^{\gamma}({\mib q})$ of the exchange integrals, where $J_1$ and $J_2$ are the intra- and inter-chain exchange interaction, respectively (we use the symbols ${\mbox{\footnotesize $/\!\!/$}}$, $\bot$ for expressing the components parallel and perpendicular to the $c$-axis).

\begin{figure}[h]
 \vspace{10pt}
  \begin{center}
    \epsfxsize=6.0cm    %  To keep the "height/width" ratio of your own figure,
    \epsfysize=4.5cm     %  please use either "epsfxsize" or "epsfysize" command.
    \epsfbox{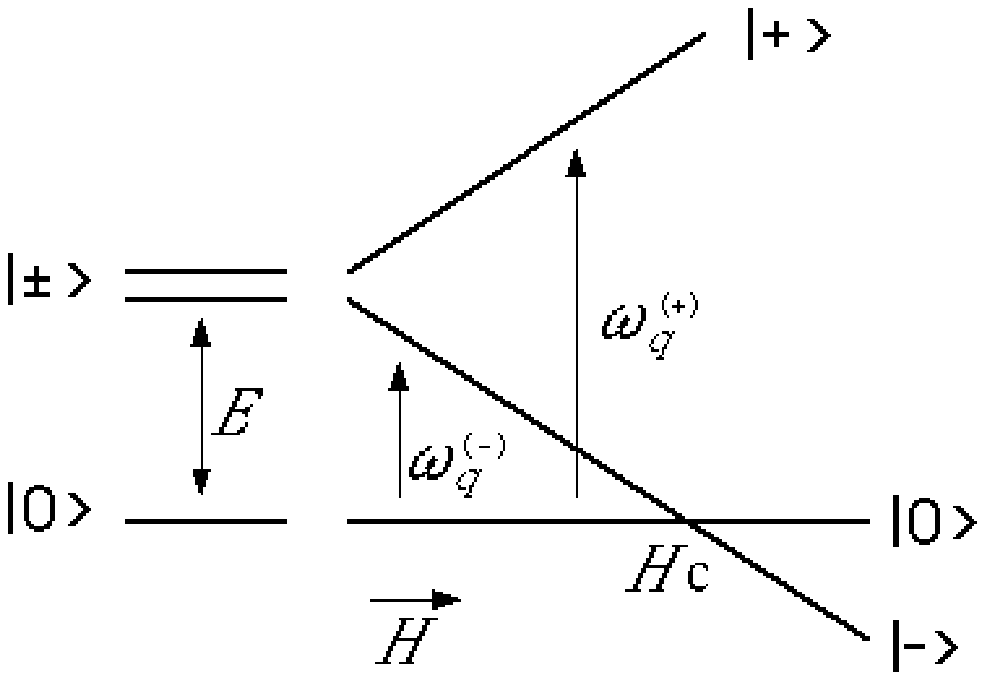}
  \end{center}
\caption{Schematic illustration of the energy level structure of the effective single-ion Hamiltonian for $H {\mbox {\scriptsize $/\!\!/$}} c$ given in eq.(\ref{effsingle}) in the text. Excitation energies represented as $\omega^{\mbox{\tiny $(+)$}}_{\mbox{\footnotesize {\boldmath $q$}}}$ and $\omega^{\mbox{\tiny $(-)$}}_{\mbox{\footnotesize {\boldmath $q$}}}$ in the general case, originate primarily from the single-ion excitations $|0\rangle \to |\pm \rangle$ for ${\mib q}=0$.}
\label{Hpara}
\end{figure}

 Figure \ref{Hpara} shows the energy level structure of the effective single-ion Hamiltonian in eq.(\ref{effsingle}) for $H {\mbox {\scriptsize $/\!\!/$}} c$. The ground state and the first excited state crosses at $H_{\rm c}$.
The energy eigenstates of $H_{i}^{\it eff}$ are given as $|1)=|0\rangle$, $|2)=|-\rangle$ and $|3)=|+\rangle$, and the corresponding eigenstates are given as $E_1=0$, $E_2=E-B$ and $E_3=E+B$, respectively (here, $|0\rangle$, $|\pm \rangle$ are the eigenstates of $S_{iz}$). The matrix representations of $S_{i\zeta}$ ($\zeta=+,-,z$) with respect to the eigenstates, $|0>$, $|->$ and $|+>$ are given as follows;

\begin{eqnarray}
S_{i+}&=&
\left(
  \begin{array}{ccc}
     0  & \sqrt{2}  &  0  \\
     0  &  0  &  0  \\
     \sqrt{2}  &  0  &  0  \\
  \end{array}
\right),
S_{i-}=
\left(
  \begin{array}{ccc}
     0  &  0  &  \sqrt{2}  \\
     \sqrt{2}  &  0  &  0  \\
     0  &  0  &  0  \\
  \end{array}
\right), \nonumber \\
S_{iz}&=&
\left(
  \begin{array}{ccc}
     0  &  0  &  0  \\
     0  &  -1  &  0  \\
     0  &  0  &  1  \\
  \end{array}
\right).
\label{matrixHpara}
\end{eqnarray}
It is noted that the magnetic excitation exists between the states connected by non zero value of the off-diagonal elements. According to the DCEFA theory, which takes into account the ${\mib q}$-dependence of magnetic excitation, the excitation energies defined as $\omega^{\mbox{\tiny $(-)$}}_{\mbox{\footnotesize {\boldmath $q$}}}$ and $\omega^{\mbox{\tiny $(+)$}}_{\mbox{\footnotesize {\boldmath $q$}}}$ in Fig.\ref{Hpara} are obtained as poles of the dynamical susceptibilities $\chi^{\mbox{\tiny $\perp$}}({\mib q}, \omega) \equiv \chi^{\mbox{\tiny $+-$}}({\mib q}, \omega)$. The derivation of the dynamical susceptibilities $\chi^{\mbox{\tiny $\zeta\eta$}}({\mib q}, \omega)$ ($\zeta, \eta= +, -, z$) are shown in Appendix A. We note that $\chi^{\mbox{\tiny $/\!\!/$}}({\mib q}, \omega) \equiv \chi^{\mbox{\tiny $zz$}}({\mib q}, \omega)$ originates from the diagonal elements of the matrix representation of $S_{iz}$, and it is derived by considering the response of the induced moment to the infinitesimal displacement of the molecular field.

 In general, the relaxation rate $T_1^{-1}$ is given by the Fourier-transform at the nuclear Larmor frequency $\omega_{\rm n}$ of the fluctuating local field transverse to the nuclear quantization axis (defined as the $Z$-axis) as follows\cite{rf:16};
\begin{eqnarray}
\frac{1}{T_1}&=&\frac{\gamma_{\rm n}^2}{2}\int_{-\infty}^{\infty}dt \langle \{ h^+(t) h^{-}(0)\}\rangle \exp(i \omega_{\rm n}t),
\label{deltah}
\end{eqnarray} 
where $\{AB\}$ denotes the symmetrized product $\frac{1}{2}(AB+BA)$. The transverse local field at the $^{133}$Cs ($^{87}$Rb) nucleus is due to the dipole field coming from the surrounding electron spins $S_{i\zeta}$ of Fe$^{2+}$ ions. Then the above equation is rewritten in the following equation,
 
\begin{eqnarray}
\frac{1}{T_1}&=&\frac{\gamma_{\rm n}^2 \mu_{\rm B}^2}{2N}\sum_{\zeta,\eta=+,-,z}\sum_{\mib q}(g_{\mbox{\tiny $\zeta$}}g_{\mbox{\tiny $\eta$}})\int_{-\infty}^{\infty}dt{\cal A}^{\mbox{\tiny $\zeta\eta$}}({\mib q})\nonumber \\
& & \times \langle \{ S_{\mbox{\footnotesize {\boldmath $q$}}}^{\mbox{\tiny $\zeta$}}(t) S_{-{\mbox{\footnotesize {\boldmath $q$}}}}^{\mbox{\tiny $\eta$}}(0)\}\rangle \exp(i\omega_{\rm n}t),
\label{paraperp}
\end{eqnarray}
where ${\cal A}^{\mbox{\tiny $\zeta\eta$}}({\mib q})$ is the geometrical factor which is calculated as the product of the dipole tensor; $S_{\mbox{\footnotesize {\boldmath $q$}}}^{\zeta}$ is the Fourier-transform of the electron spin operator $S_{i\zeta}$ with respect to the wave vector ${\mib q}$.

 As we described in the previous paragraph, we obtain the eigenfrequency $\omega^{({\mbox{\tiny $\xi$}})}_{\mbox{\footnotesize {\boldmath $q$}}}$ ($\xi=+,-$) of the singlet ground state system. Here, we extract explicitly from the time dependence of the spin operator the part related to $\omega^{({\mbox{\tiny $\xi$}})}_{\mbox{\footnotesize {\boldmath $q$}}}$. We use the fluctuation-dissipation theorem between the spin correlation function and the relaxation function as \cite{rf:17};
\begin{equation}
\int^{\infty}_{-\infty}dt e^{i\omega t}\langle \{\delta S_{\mbox{\footnotesize {\boldmath $q$}}}^{\mbox{\tiny $\zeta$}}(t) \delta S_{-{\mbox{\footnotesize {\boldmath $q$}}}}^{\mbox{\tiny $\eta$}}(0)\} \rangle = E_{\beta}(\omega)R^{\mbox{\tiny $\zeta\eta$}}_{\mbox{\footnotesize {\boldmath $q$}}}(\omega),
\label{sanitsu}
\end{equation}
with 
\begin{equation}
E_{\beta}(\omega)= \frac{\hbar\omega}{2}\coth(\frac{\hbar\omega}{2k_{\mbox{\tiny B}}T}),
\end{equation}
 where $\omega$ is taken to be $\omega^{({\mbox{\tiny $\xi$}})}_{\mbox{\footnotesize {\boldmath $q$}}}$ or $\omega_{\rm n}$ according to the relevant components, and $R^{\mbox{\tiny $\zeta\eta$}}_{\mbox{\footnotesize {\boldmath $q$}}}(\omega)$ is the Fourier transform of the relaxation function $R^{\mbox{\tiny $\zeta\eta$}}_{\mbox{\footnotesize {\boldmath $q$}}}(t)$. We here assume that $R^{\mbox{\tiny $\zeta\eta$}}_{\mbox{\footnotesize {\boldmath $q$}}}(t)$ in eq.(\ref{sanitsu}) is described in terms of ${\mib q}$-dependent static susceptibility $\chi^{\mbox{\tiny $\zeta\eta$}}({\mib q})$ and normalized relaxation function $f_{\mbox{\footnotesize {\boldmath $q$}}}^{\mbox{\tiny $\zeta\eta$}}(t)$ as
\begin{eqnarray}
R_{\mbox{\footnotesize {\boldmath $q$}}}^{\mbox{\tiny $\zeta\eta$}}(t)=\chi^{\mbox{\tiny $\zeta\eta$}}({\mib q})f_{\mbox{\footnotesize {\boldmath $q$}}}^{\mbox{\tiny $\zeta\eta$}}(t).
\label{kanwa}
\end{eqnarray}
The static ${\mib q}$-dependent susceptibilities $\chi^{\mbox{\tiny $\zeta\eta$}}({\mib q})$ are evaluated by putting $\omega=0$ in the dynamical susceptibilities $\chi^{\mbox{\tiny $\zeta\eta$}}({\mib q}, \omega)$. Then, the dynamical nature in $\chi^{\mbox{\tiny $\zeta\eta$}}({\mib q}, \omega)$ is transferred to the normalized relaxation function, and the problem is reduced to the evaluation of $f_{\mbox{\footnotesize {\boldmath $q$}}}^{\mbox{\tiny $\zeta\eta$}}(t)$.

 Now we look at the normalized relaxation function in eq.(\ref{kanwa}). As shown in the previous paragraph, the excited state of singlet ground state is charactrized by the excitation energy $\omega^{\mbox{\tiny $(\xi)$}}_{\mbox{\footnotesize {\boldmath $q$}}}$. Then, when the damping effect is neglected, the time evolution of the excited state is described as $\exp(i\omega^{\mbox{\tiny $(\xi)$}}_{\mbox{\footnotesize {\boldmath $q$}}}t)$. Let us take into account the damping effect which is expressed by the exponential function. 
 
 As for the fluctuation transverse to the external field, we define $f_{\mbox{\footnotesize {\boldmath $q$}}}^{\mbox{\tiny $\perp$}}(t)$ and disregard the effect of the upper branch $\omega^{\mbox{\tiny $(+)$}}_{\mbox{\footnotesize {\boldmath $q$}}}$. Treating the time evolution of the lower branch $\omega^{\mbox{\tiny $(-)$}}_{\mbox{\footnotesize {\boldmath $q$}}}$ which is effective to the nuclear magnetic relaxation, the normalized relaxation function may be given as follows;
\begin{equation}
f_{\mbox{\footnotesize {\boldmath $q$}}}^{\mbox{\tiny $\perp$}}(t)=\exp(i\omega_{\mbox{\footnotesize {\boldmath $q$}}}^{\mbox{\tiny $(-)$}}t-{\it \Gamma_{\mbox{\footnotesize {\boldmath $q$}}}^{\mbox{\tiny $\perp$}}}t).
\label{fperpg}
\end{equation}

Physical meaning of ${\it \Gamma^{\mbox{\tiny $\perp$}}_{\mbox{\footnotesize {\boldmath $q$}}}}$ can be considered to be the lifetime of magnetic excitation $\omega^{\mbox{\tiny $(-)$}}_{\mbox{\footnotesize {\boldmath $q$}}}$, and broadening of the spectrum occurs due to ${\it \Gamma^{\mbox{\tiny $\perp$}}_{\mbox{\footnotesize {\boldmath $q$}}}}$. The concept of lifetime is not considered within the frame of the DCEFA theory; however, when there is interaction between the excitations, or the higher order terms of interactions are included, then the magnetic excitation may have lifetime whose value is given by the inverse of ${\it \Gamma^{\mbox{\tiny $\perp$}}_{\mbox{\footnotesize {\boldmath $q$}}}}$. 

 As for $f^{\mbox{\tiny $/\!\!/$}}_{\mbox{\footnotesize {\boldmath $q$}}}(t)$, considering that this is the fluctuation of the induced moment along the external field, we may put 
\begin{equation}
f^{\mbox{\tiny $/\!\!/$}}_{\mbox{\footnotesize {\boldmath $q$}}}(t) \equiv f^{\mbox{\tiny $zz$}}_{\mbox{\footnotesize {\boldmath $q$}}}(t)=\exp(-{\it \Gamma^{\mbox{\tiny $/\!\!/$}}_{\mbox{\footnotesize {\boldmath $q$}}}}t). 
\label{fparaqzz}
\end{equation}
 Arrangement of the induced moment might be disordered by the propagation of magnetic excitation (or the transverse fluctuation), then this could be the physical origin of ${\it \Gamma^{\mbox{\tiny $/\!\!/$}}_{\mbox{\footnotesize {\boldmath $q$}}}}$.
  
\begin{figure}[h]
 \vspace{10pt}
  \begin{center}
    \epsfxsize=8.0cm    %  To keep the "height/width" ratio of your own figure,
    \epsfysize=4.8cm     %  please use either "epsfxsize" or "epsfysize" command.
    \epsfbox{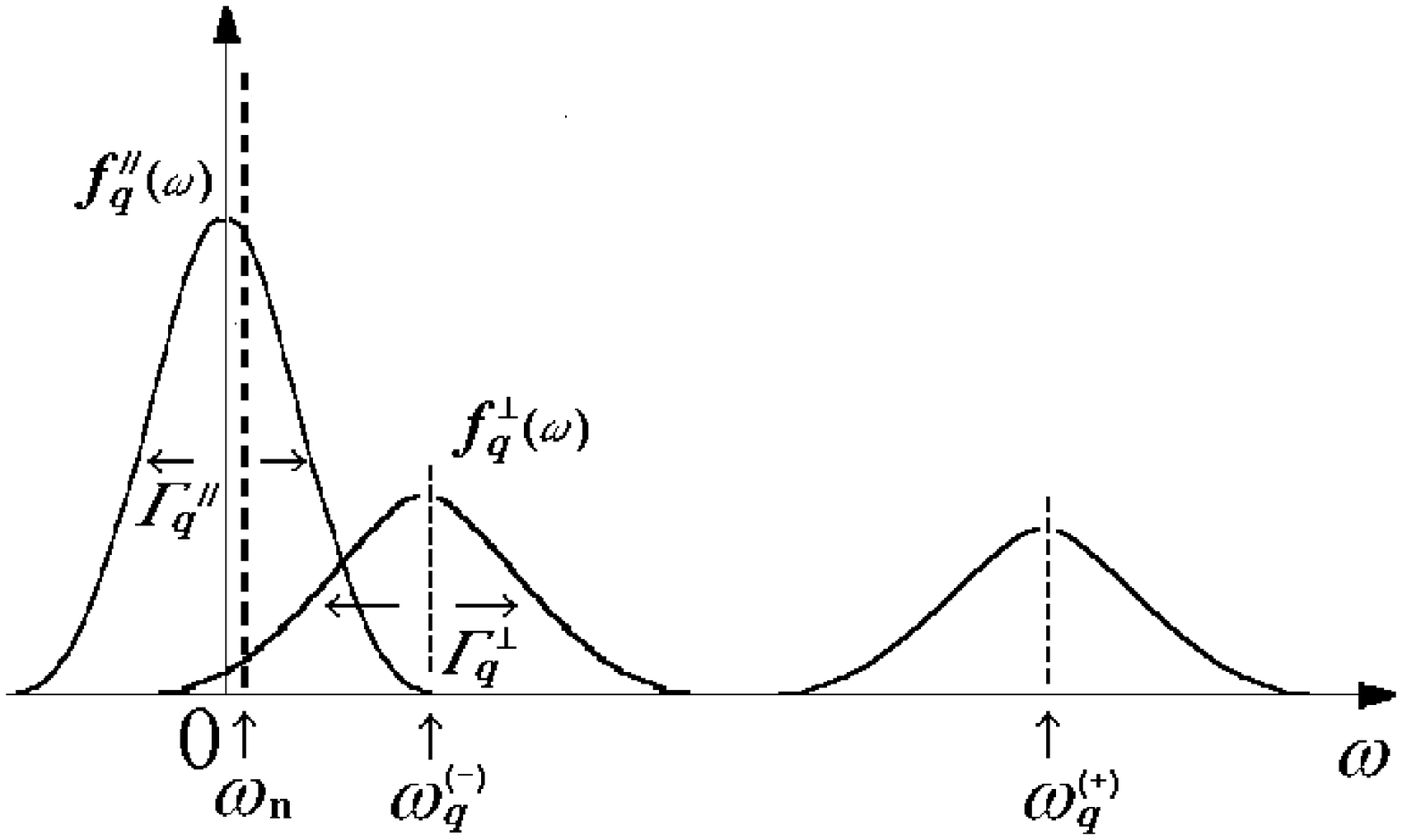}
  \end{center}
\caption{ An illustration of the frequency dependence of $f^{\mbox{\tiny $\perp$}}_{\mbox{\footnotesize {\boldmath $q$}}}(\omega)$ and $f^{\mbox{\tiny $/\!\!/$}}_{\mbox{\footnotesize {\boldmath $q$}}}(\omega)$.}
\label{freq}
\end{figure}

 Here we define $f^{\mbox{\tiny $\gamma$}}_{\mbox{\footnotesize {\boldmath $q$}}}(\omega)$ as the Fourier-transform of $f^{\mbox{\tiny $\gamma$}}_{\mbox{\footnotesize {\boldmath $q$}}}(t)$ with respect to $\omega$. Figure \ref{freq} shows an illustration of the frequency dependence of $f^{\mbox{\tiny $\perp$}}_{\mbox{\footnotesize {\boldmath $q$}}}(\omega)$ and $f^{\mbox{\tiny $/\!\!/$}}_{\mbox{\footnotesize {\boldmath $q$}}}(\omega)$ near $\omega=0$. $f^{\mbox{\tiny $/\!\!/$}}_{\mbox{\footnotesize {\boldmath $q$}}}(\omega)$ and $f^{\mbox{\tiny $\perp$}}_{\mbox{\footnotesize {\boldmath $q$}}}(\omega)$ have their centers at zero and $\omega^{\mbox{\tiny $(-)$}}_{\mbox{\footnotesize {\boldmath $q$}}}$, and broadening of the spectra are characterized by ${\it \Gamma^{\mbox{\tiny $/\!\!/$}}_{\mbox{\footnotesize {\boldmath $q$}}}}$ and ${\it \Gamma^{\mbox{\tiny $\perp$}}_{\mbox{\footnotesize {\boldmath $q$}}}}$. 
 As shown in Fig.\ref{freq}, the relaxation rate will be explained in terms of two different kinds of electron spin fluctuations, and the ratio of each contribution will be determined by the values of ${\it \Gamma^{\mbox{\tiny $\perp$}}_{\mbox{\footnotesize {\boldmath $q$}}}}$ and ${\it \Gamma^{\mbox{\tiny $/\!\!/$}}_{\mbox{\footnotesize {\boldmath $q$}}}}$ (the effect of the excitation of $\omega^{\mbox{\tiny $(+)$}}_{\mbox{\footnotesize {\boldmath $q$}}}$ mode can be ignored). If there is no broadening of spectrum due to ${\it \Gamma^{\mbox{\tiny $\gamma$}}_{\mbox{\footnotesize {\boldmath $q$}}}}$, then $f^{\mbox{\tiny $\gamma$}}_{\mbox{\footnotesize {\boldmath $q$}}}(\omega)$ do not have frequency of $\omega_n$ component, and this fluctuation will not work as a thermal reservoir of the nuclear spin relaxation. 

 Referring to the above considerations, the theoretical equation of $T_1^{-1}$ is given as $T_1^{-1}=(T_1^{-1})^{\mbox{\tiny $/\!\!/$}}+(T_1^{-1})^{\mbox{\tiny $\perp$}}$ with
\begin{equation}
(T_1^{-1})^{\mbox{\tiny $/\!\!/$}}=\frac{\gamma_{\rm n}^2}{2} \frac{1}{N}\sum_{{\mib q}} (g_{\mbox{\tiny $/\!\!/$}} \mu_{\mbox{\tiny B}})^2{\cal A}^{\mbox{\tiny $/\!\!/$}}({\mib q}) k_{\mbox{\tiny B}} T\chi^{\mbox{\tiny $/\!\!/$}}({\mib q})\frac{2}{{\it \Gamma_{\mbox{\footnotesize {\boldmath $q$}}}^{\mbox{\tiny $/\!\!/$}}}}, \label{Tz}
\end{equation}
\begin{eqnarray}
(T_1^{-1})^{\mbox{\tiny $\perp$}} &=& \frac{\gamma_{\rm n}^2}{2}\frac{1}{N} \sum_{{\mib q}} (g_{\mbox{\tiny $\perp$}} \mu_{\mbox{\tiny B}})^2 {\cal A}^{\mbox{\tiny $\perp$}}({\mib q})\frac{\hbar\omega}{2}\hspace{25mm} \nonumber \\
& & \hspace{3mm} \times\coth(\frac{\hbar\omega}{2k_{\mbox{\tiny B}}T}) \chi^{\mbox{\tiny $\perp$}}({\mib q})\frac{2{\it \Gamma_{\mbox{\footnotesize {\boldmath $q$}}}^{\mbox{\tiny $\perp$}}}}{{\it \Gamma_{\mbox{\footnotesize {\boldmath $q$}}}^{\mbox{\tiny $\perp$}}}^2+ (\omega)^2},
\label{Txy}
\end{eqnarray}
 where ${\it \Gamma_{\mbox{\footnotesize {\boldmath $q$}}}^{\mbox{\tiny $/\!\!/$}}} \gg \omega_{\rm n}$ is assumed for eq.(\ref{Tz}), and $\omega=\omega_{\mbox{\footnotesize {\boldmath $q$}}}^{\mbox{\tiny $(-)$}}+\omega_{\rm n}$ in eq.(\ref{Txy}); considering that $\hbar\omega_{\rm n} \ll k_{\mbox{\tiny B}}T$ for the relevant temperatures, we put $E_{\beta}(\omega_{\rm n}) \approx k_{\mbox{\tiny B}} T$ for eq.(\ref{Tz}). The expressions of the geometrical factors ${\cal A}^{\mbox{\tiny $\gamma$}}({\mib q})$ are shown in Appendix B.

\subsection{$H \bot c$} 
 Next we consider the case for $H {\mbox {\scriptsize $/\!\!/$}} x$ (crystalline $a$-axis), which corresponds to the present experimental condition. The effective single-ion Hamiltonian is expressed as \cite{rf:22}
\begin{equation}
H_{i}^{\it eff}=ES_{iz}^2 +B^{\mbox{\tiny $\perp$}}S_{ix},
\label{effsingleperp}
\end{equation}
with $B^{\mbox{\tiny $\perp$}}=g_{\mbox{\tiny $\perp$}}\mu_BH + 2J^{\mbox{\tiny $\perp$}}(0)(1-\alpha)\langle S_x \rangle$. Here $\langle S_x \rangle $ is the thermal average of the induced moment. 

\begin{figure}[h]
 \vspace{10pt}
  \begin{center}
    \epsfxsize=6.0cm    %  To keep the "height/width" ratio of your own figure,
    \epsfysize=4.5cm     %  please use either "epsfxsize" or "epsfysize" command.
    \epsfbox{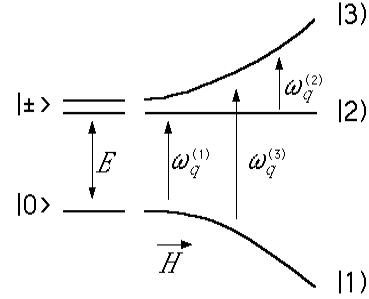}
  \end{center}
\caption{Schematic illustration of the energy level structure of the effective single-ion Hamiltonian for $H\bot c$ ($H {\mbox {\scriptsize $/\!\!/$}} x$). Excitation energies represented as $\omega^{\mbox{\tiny $(1)$}}_{\mib q}$, $\omega^{\mbox{\tiny $(2)$}}_{\mib q}$ and $\omega^{\mbox{\tiny $(3)$}}_{\mib q}$ in the general case, originate primarily from the single-ion excitations $|1)\to|2)$, $|2)\to|3)$ and $|1)\to|3)$ for ${\mib q}=0$.}
\label{Hperpc}
\end{figure}

  Figure \ref{Hperpc} shows the energy level structure of the effective single-ion Hamiltonian. 
The gap energy between the ground state and the first excited state increases with increasing external field. The eigenstates of $H_{i}^{\it eff}$, which are given by the linear combinations of $|\pm \rangle $ and $|0 \rangle $, and the corresponding energy eigenvalues are given as follows,
\begin{eqnarray}
|1)&=&d_{\mbox{\tiny $1$}}/\sqrt{2}(|+ \rangle +|- \rangle)+\sqrt{2}c_{\mbox{\tiny $1$}}|0 \rangle; \hspace{3mm} E_1=1/2(E-W), \nonumber \\
  & &  \\
|2)&=&(|+ \rangle -|- \rangle)/\sqrt{2}; \hspace{3mm} E_2=E, \\
|3)&=&c_{\mbox{\tiny $2$}}(|+ \rangle +|- \rangle)-d_{\mbox{\tiny $2$}}|0 \rangle; \hspace{3mm} E_3=1/2(E+W), 
\end{eqnarray}
where $W=(E^2+4B^2)^{1/2}$, $F=\{(E+W)^2+4B^2\}^{1/2}$, $c_{\mbox{\tiny $1$}}=\sqrt{2}B/F$, $d_{\mbox{\tiny $1$}}=(E+W)/F$, $c_{\mbox{\tiny $2$}}=d_{\mbox{\tiny $1$}}/\sqrt{2}$ and $d_{\mbox{\tiny $2$}}=\sqrt{2}c_{\mbox{\tiny $1$}}$.
 The matrix representations of $S_{i\mu}$ ($\mu=x, y, z$) with respect to the eigenstates, $|1)$, $|2)$ and $|3)$ are given as follows;
\begin{eqnarray}
S_{ix}&=&
\left(
  \begin{array}{ccc}
    2\sqrt{2}d_{\mbox{\tiny $1$}}c_{\mbox{\tiny $1$}}   &  0  &  \sqrt{2}(d_{\mbox{\tiny $1$}}c_{\mbox{\tiny $2$}}-d_{\mbox{\tiny $2$}}c_{\mbox{\tiny $1$}})  \\
     0  &  0  &  0  \\
    \sqrt{2}(d_{\mbox{\tiny $1$}}c_{\mbox{\tiny $2$}}-d_{\mbox{\tiny $2$}}c_{\mbox{\tiny $1$}})   &  0  &   -2\sqrt{2}d_{\mbox{\tiny $1$}}c_{\mbox{\tiny $1$}}  \\
  \end{array}
\right), \nonumber \\
S_{iy}&=&
\left(
  \begin{array}{ccc}
     0  &  id_{\mbox{\tiny $1$}}  & 0 \\
     -id_{\mbox{\tiny $1$}}  &  0  &  id_{\mbox{\tiny $2$}}  \\
     0  &  -id_{\mbox{\tiny $2$}} &  0  \\
  \end{array}
\right), \nonumber \\
S_{iz}&=&
\left(
  \begin{array}{ccc}
     0  &  \sqrt{2}c_{\mbox{\tiny $1$}}  &  0  \\
     \sqrt{2}c_{\mbox{\tiny $1$}}  &  0  &  \sqrt{2}c_{\mbox{\tiny $2$}}  \\
     0  &  \sqrt{2}c_{\mbox{\tiny $2$}}  &  0  \\
  \end{array}
\right).
\label{Sxmatrix}
\end{eqnarray}
According to the DCEFA theory, excitation energies are defined as $\omega^{\mbox{\tiny $(1)$}}_{\mbox{\footnotesize {\boldmath $q$}}}$, $\omega^{\mbox{\tiny $(2)$}}_{\mbox{\footnotesize {\boldmath $q$}}}$ and $\omega^{\mbox{\tiny $(3)$}}_{\mbox{\footnotesize {\boldmath $q$}}}$; $\omega^{\mbox{\tiny $(1)$}}_{\mbox{\footnotesize {\boldmath $q$}}}$ and $\omega^{\mbox{\tiny $(2)$}}_{\mbox{\footnotesize {\boldmath $q$}}}$ are obtained as poles of $\chi^{yy}({\mib q}, \omega)$, $\chi^{yz}({\mib q}, \omega)$ and $\chi^{zz}({\mib q}, \omega)$, and $\omega^{\mbox{\tiny $(3)$}}_{\mbox{\footnotesize {\boldmath $q$}}}$ is obtained as the pole of $\chi^{xx}({\mib q},\omega)$. The expressions of the dynamical susceptibilities $\chi^{\mu\nu}({\mib q}, \omega)$ ($\mu, \nu=x, y, z$) are given in Appendix A. The terms of $\chi^{yy}({\mib q},\omega)$ and $\chi^{zz}({\mib q},\omega)$ can be expressed as a sum of two terms which have poles at $\omega^{\mbox{\tiny $(1)$}}_{\mbox{\footnotesize {\boldmath $q$}}}$ and $\omega^{\mbox{\tiny $(2)$}}_{\mbox{\footnotesize {\boldmath $q$}}}$ respectively, as shown in eq.(\ref{twopole}). As for $\chi^{xx}({\mib q},\omega)$, two expressions are given for $\phi^{xx}(\omega)$ according to the case of $\omega = 0$ and $\omega \neq 0$ as shown in eq.(\ref{phixx}). The former case results from the diagonal elements of $S_{ix}$ in eq.(\ref{Sxmatrix}); we define $\chi^{xx}_{\omega_0}({\mib q}, \omega)$, which is derived by substituting the expression of $\phi^{xx}(\omega=0)$ into eq.(\ref{chixxqomega}). The latter case results from the off-diagonal elements of $S_{ix}$ in eq.(\ref{Sxmatrix}), and $\omega^{\mbox{\tiny $(3)$}}_{\mbox{\footnotesize {\boldmath $q$}}}$ is obtained as the pole of $\chi^{xx}({\mib q},\omega)$, which is derived by substituting the expression of $\phi^{xx}(\omega \neq 0)$ into eq.(\ref{chixxqomega}).
 
 As for the normalized relaxation function, we use the relation in eq.(\ref{twopole}) and treat the time evolution of the transverse spin fluctuation  separately with respect to $\omega^{\mbox{\tiny $(1)$}}_{\mbox{\footnotesize {\boldmath $q$}}}$ and $\omega^{\mbox{\tiny $(2)$}}_{\mbox{\footnotesize {\boldmath $q$}}}$. Then the normalized relaxation function for the dynamical susceptibilities $\chi^{\mu\nu}_{\omega_1}({\mib q},\omega)$ and $\chi^{\mu\nu}_{\omega_2}({\mib q},\omega)$ ($\mu,\nu=y,z$) may be given as follows;
\begin{eqnarray}
f_{\mbox{\footnotesize {\boldmath $q$}}}^{\mbox{\tiny $(1)$}}(t)&=&\exp(i\omega_{\mbox{\footnotesize {\boldmath $q$}}}^{\mbox{\tiny $(1)$}}t-{\it \Gamma_{\mbox{\footnotesize {\boldmath $q$}}}^{\mbox{\tiny $(1)$}}}t), \\
f_{\mbox{\footnotesize {\boldmath $q$}}}^{\mbox{\tiny $(2)$}}(t)&=&\exp(i\omega_{\mbox{\footnotesize {\boldmath $q$}}}^{\mbox{\tiny $(2)$}}t-{\it \Gamma_{\mbox{\footnotesize {\boldmath $q$}}}^{\mbox{\tiny $(2)$}}}t).
\label{f12}
\end{eqnarray}

 As for the fluctuation $\chi^{xx}({\mib q},\omega)$ which is along the external field, we do not take into account of the effect of the $\omega^{\mbox{\tiny $(3)$}}_{\mbox{\footnotesize {\boldmath $q$}}}$ mode; because it does not show softening and the excitation energy increases largely by applying an external field. On the other hand, we take into account of the longitudinal fluctuation of the induced moment, and the normalized relaxation function for this fluctuation is defined as 
\begin{equation}
f^{xx}_{\mbox{\footnotesize {\boldmath $q$}}}(t)=\exp(-{\it \Gamma^{\mbox{\tiny $(0)$}}_{\mbox{\footnotesize {\boldmath $q$}}}}t), 
\label{fparaqxx}
\end{equation}
where suffix (0) is used to distinguish the damping constant of the longitudinal fluctuation for $H \bot c$ from that of $H {\mbox {\scriptsize $/\!\!/$}} c$. 

 Then the theoretical equation of $T_1^{-1}$ is given as $T_1^{-1}=(T_1^{-1})_{\mbox{\tiny $1$}}+(T_1^{-1})_{\mbox{\tiny $2$}}+(T_1^{-1})_{\mbox{\tiny $0$}}$ with
\begin{eqnarray}
(T_1^{-1})_{\mbox{\tiny $1$}}&=&\frac{\gamma_{\rm n}^2}{2} \frac{1}{N}\sum_{{\mib q}} \frac{\hbar \omega }{2}\coth (\frac{\hbar \omega}{2})\{(g_{\mbox{\tiny $\perp$}} \mu_{\mbox{\tiny B}})^2{\cal A}^{yy}({\mib q})\chi_{\omega_1}^{yy}({\mib q}) \nonumber \\
&+&(g_{\mbox{\tiny $/\!\!/$}} \mu_{\mbox{\tiny B}})^2{\cal A}^{zz}({\mib q})\chi_{\omega_1}^{zz}({\mib q})\}\frac{2{\it \Gamma^{\mbox{\tiny $(1)$}}_{\mbox{\footnotesize {\boldmath $q$}}}}}{({\it \Gamma^{\mbox{\tiny $(1)$}}_{\mbox{\footnotesize {\boldmath $q$}}}})^2+(\omega )^2},
\label{T11}
\end{eqnarray}
\begin{eqnarray}
(T_1^{-1})_{\mbox{\tiny $2$}}&=&\frac{\gamma_{\rm n}^2}{2} \frac{1}{N}\sum_{{\mib q}} \frac{\hbar \omega }{2}\coth (\frac{\hbar \omega}{2})\{(g{\mbox{\tiny $\perp$}} \mu_{\mbox{\tiny B}})^2{\cal A}^{yy}({\mib q})\chi_{\omega_2}^{yy}({\mib q}) \nonumber \\
&+&(g_{\mbox{\tiny $/\!\!/$}} \mu_{\mbox{\tiny B}})^2{\cal A}^{zz}({\mib q})\chi_{\omega_2}^{zz}({\mib q})\}\frac{2{\it \Gamma^{\mbox{\tiny $(2)$}}_{\mbox{\footnotesize {\boldmath $q$}}}}}{({\it \Gamma^{\mbox{\tiny $(2)$}}_{\mbox{\footnotesize {\boldmath $q$}}}})^2+(\omega )^2},
\label{T12} \\
(T_1^{-1})_{\mbox{\tiny $0$}}&=&\frac{\gamma_{\rm n}^2}{2} \frac{1}{N}\sum_{{\mib q}} k_{\mbox{\tiny B}} T (g_{\mbox{\tiny $\perp$}} \mu_{\mbox{\tiny B}})^2{\cal A}^{xx}({\mib q}) \chi_{\omega_0}^{xx}({\mib q})\frac{2}{{\it \Gamma^{\mbox{\tiny $(0)$}}_{\mbox{\footnotesize {\boldmath $q$}}}}},
\label{T0}
\end{eqnarray}
where $\omega =\omega^{\mbox{\tiny $(1)$}}_{\mbox{\footnotesize {\boldmath $q$}}}+\omega_{\rm n}$ for eq.(\ref{T11}), $\omega =\omega^{\mbox{\tiny $(2)$}}_{\mbox{\footnotesize {\boldmath $q$}}}+\omega_{\rm n}$ for eq.(\ref{T12}) and ${\it \Gamma_{\mbox{\footnotesize {\boldmath $q$}}}^{\mbox{\tiny $(0)$}}} \gg \omega_{\rm n}$ is assumed for eq.(\ref{T0}). The expressions of ${\cal A}^{\mu\nu}({\mib q})$ are shown in Appendix B. From the relation ${\cal A}^{yz}({\mib q})={\cal A}^{zy}({\mib q})$, and $\chi^{yz}({\mib q},\omega)=-\chi^{zy}({\mib q},\omega)$ in eq.(\ref{chiyz}), we obtain the following equation,
\begin{equation}
{\cal A}^{yz}({\mib q})\cdot \chi^{yz}({\mib q},\omega)+{\cal A}^{zy}({\mib q})\cdot \chi^{zy}({\mib q},\omega)=0,
\end{equation}
 thus $\chi^{zy}({\mib q},\omega)$ and $\chi^{yz}({\mib q},\omega)$ do not have any contribution to the relaxation rate. 

\section{Analysis and Discussion}
 In the previous section, the expressions of the relaxation rate have been shown as the sum of the contribution of different magnetic modes. We introduced the damping constants for each magnetic excitation in the singlet ground state system. It is difficult to evaluate these factors, theoretically; then, we determine the values of damping constants by the fitting of the theoretical curve to the experimental results of $T_1^{-1}$. The numerical values of Hamiltonian parameters for CFC and RFC are shown in Tabel \ref{hamiltonian para}, which have already been evaluated by Suzuki and Makino from the experimental results of ESR and neutron-scattering\cite{rf:2}.
% 表の挿入
\begin{table}[htbp]
 \caption{The Hamiltonian parameters of CFC and RFC evaluated by Suzuki and Makino\cite{rf:2}.}% {}内に表題を書く
 \begin{center}
  \begin{tabular}{@{\hspace{\tabcolsep}\extracolsep{\fill}}ccc}
    \hline
       &CsFeCl$_3$ & RbFeCl$_3$   \\
    \hline
     $g_{\mbox{\tiny $/\!\!/$}}$  &  2.54  &  2.54  \\
     $g_{\mbox{\tiny $\perp$}}$  &  3.84  &  3.84  \\
     $D/k_{\mbox{\tiny B}}$ (K) &  20.1  &  21.3  \\
     $J_1^{\mbox{\tiny $\perp$}}/k_{\mbox{\tiny B}}$ (K) &  3.72  &  5.05  \\
     $J_2^{\mbox{\tiny $\perp$}}/k_{\mbox{\tiny B}}$ (K) &  -0.201  &  -0.429  \\
     $|J_2/J_1|$  &  0.055  &  0.085  \\
     $J^{\mbox{\tiny $/\!\!/$}}/J^{\mbox{\tiny $\perp$}}$  &  1.0  & 1.3   \\
    \hline
  \end{tabular}
 \end{center}
 \label{hamiltonian para}
\end{table}

\subsection{$H {\mbox {\scriptsize $/\!\!/$}} c$}
First, we consider the case of $H {\mbox {\scriptsize $/\!\!/$}} c$ in CFC.\hspace{3mm}
\begin{figure}[htbp]
  \begin{center}
    \epsfxsize=8.0cm    %  To keep the "height/width" ratio of your own figure,
    \epsfysize=6.2cm     %  please use either "epsfxsize" or "epsfysize" command.
    \epsfbox{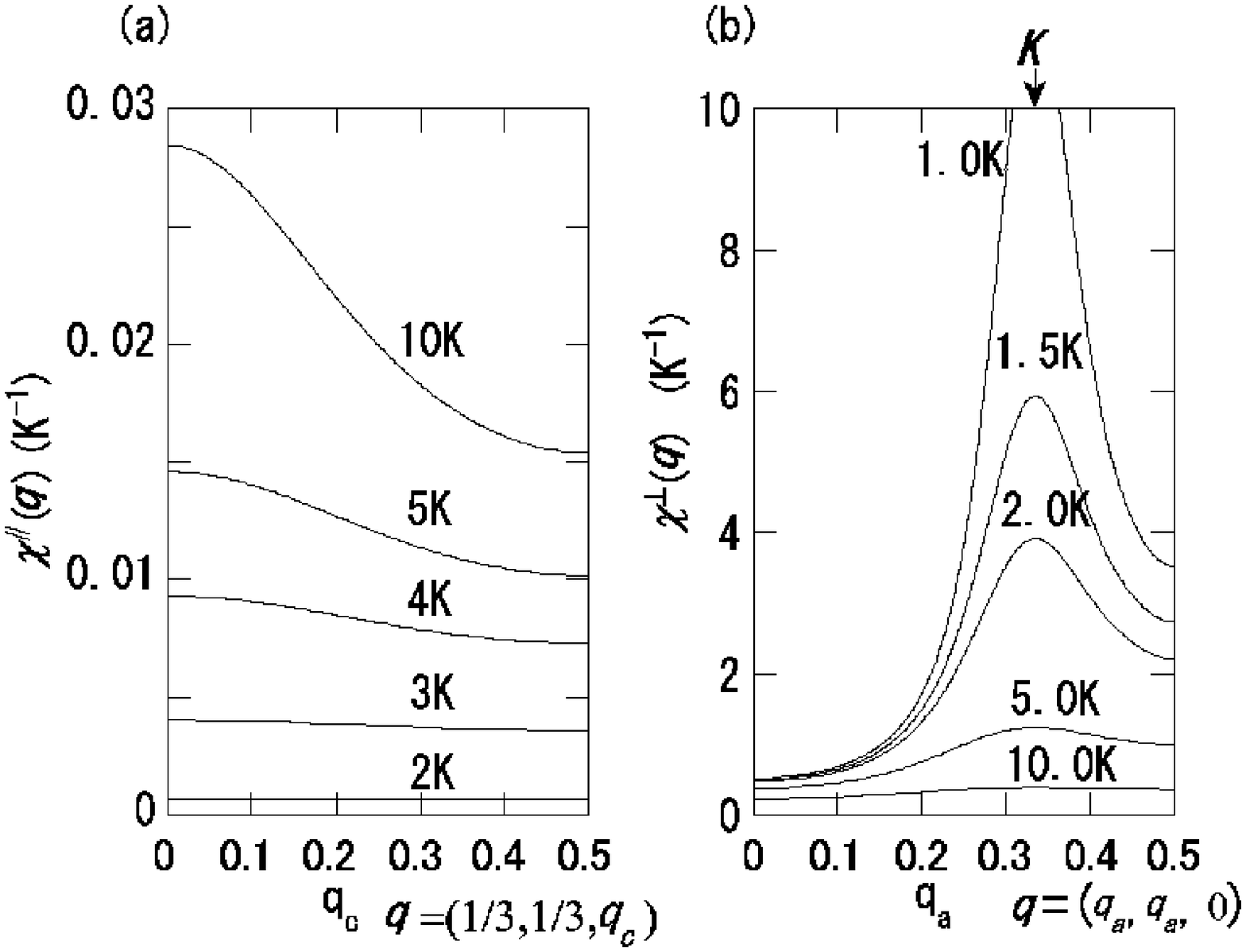}
  \end{center}
\caption{(a) Temperature dependence of $\chi^{\mbox{\tiny $/\!\!/$}}({\mib q})$ along ${\mib q}=(1/3,1/3,q_c)$ and (b) $\chi^{\mbox{\tiny $\perp$}}({\mib q})$ along ${\mib q}=(q,q,0)$, calculated for CFC at $H$=5.0 T on the basis of the DCEFA.}
\label{kaixxparaq}
\end{figure}
\begin{figure}[b]
  \begin{center}
    \epsfxsize=8.0cm    %  To keep the "height/width" ratio of your own figure,
    \epsfysize=6.2cm     %  please use either "epsfxsize" or "epsfysize" command.
    \epsfbox{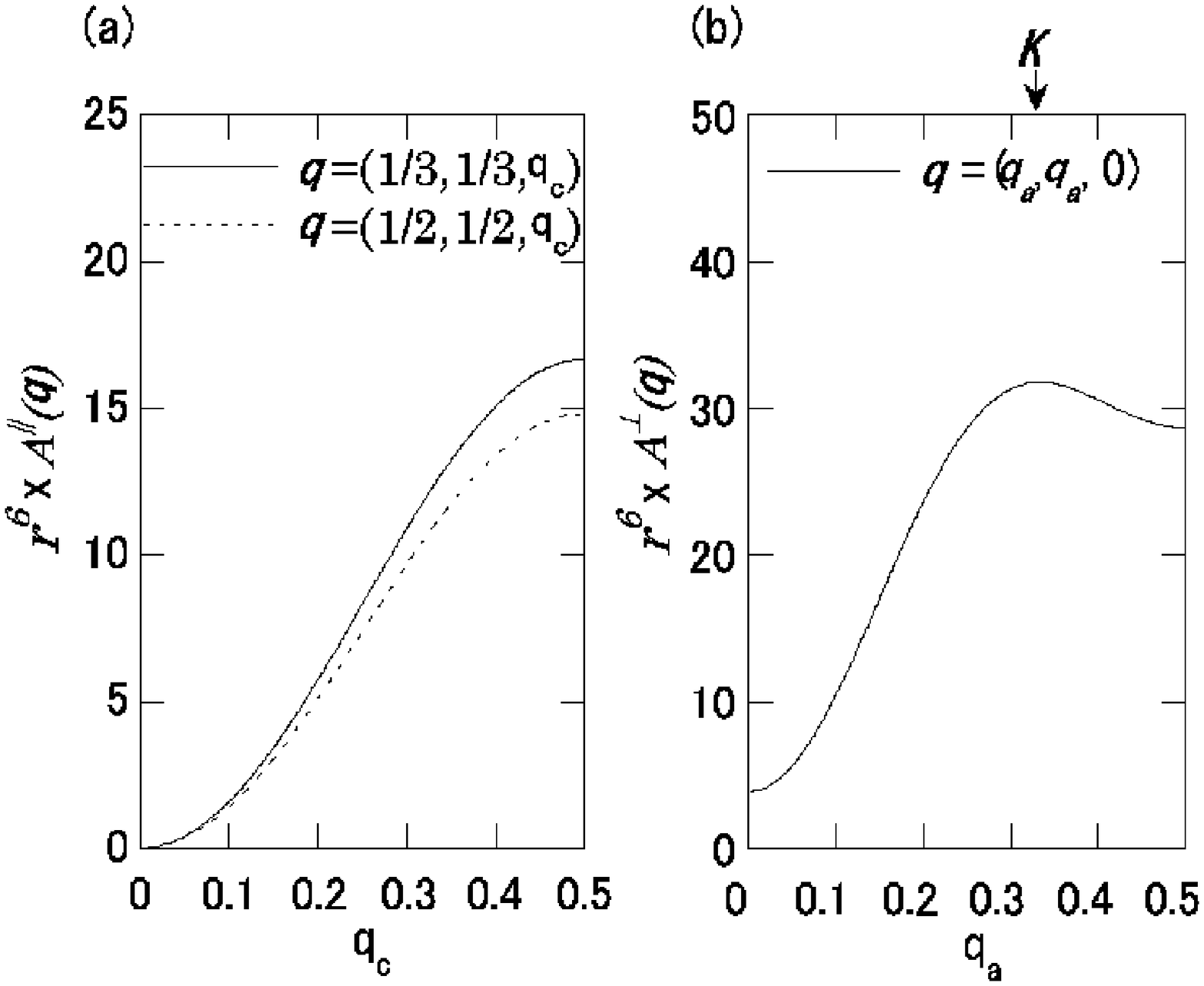}
  \end{center}
\caption{(a) Geometrical factor $r^6 \times {\cal A}^{\mbox{\tiny $/\!\!/$}}({\mib q})$ along ${\mib q}=(1/3,1/3,q_c)$ and ${\mib q}=(1/2,1/2,q_c)$, and (b) $r^6 \times{\cal A}^{\mbox{\tiny $\perp$}}({\mib q})$ along ${\mib q}=(q,q,0)$ where ${\it r}$ is the distance between the $^{133}$Cs nucleus and the nearest Fe$^{2+}$ ions (${\it r}=4.4 \times 10^{-8}$ cm).}
\label{Geoparaq}
\end{figure}
\begin{figure}[b]
  \begin{center}
    \epsfxsize=8.0cm    %  To keep the "height/width" ratio of your own figure,
    \epsfysize=9.0cm     %  please use either "epsfxsize" or "epsfysize" command.
    \epsfbox{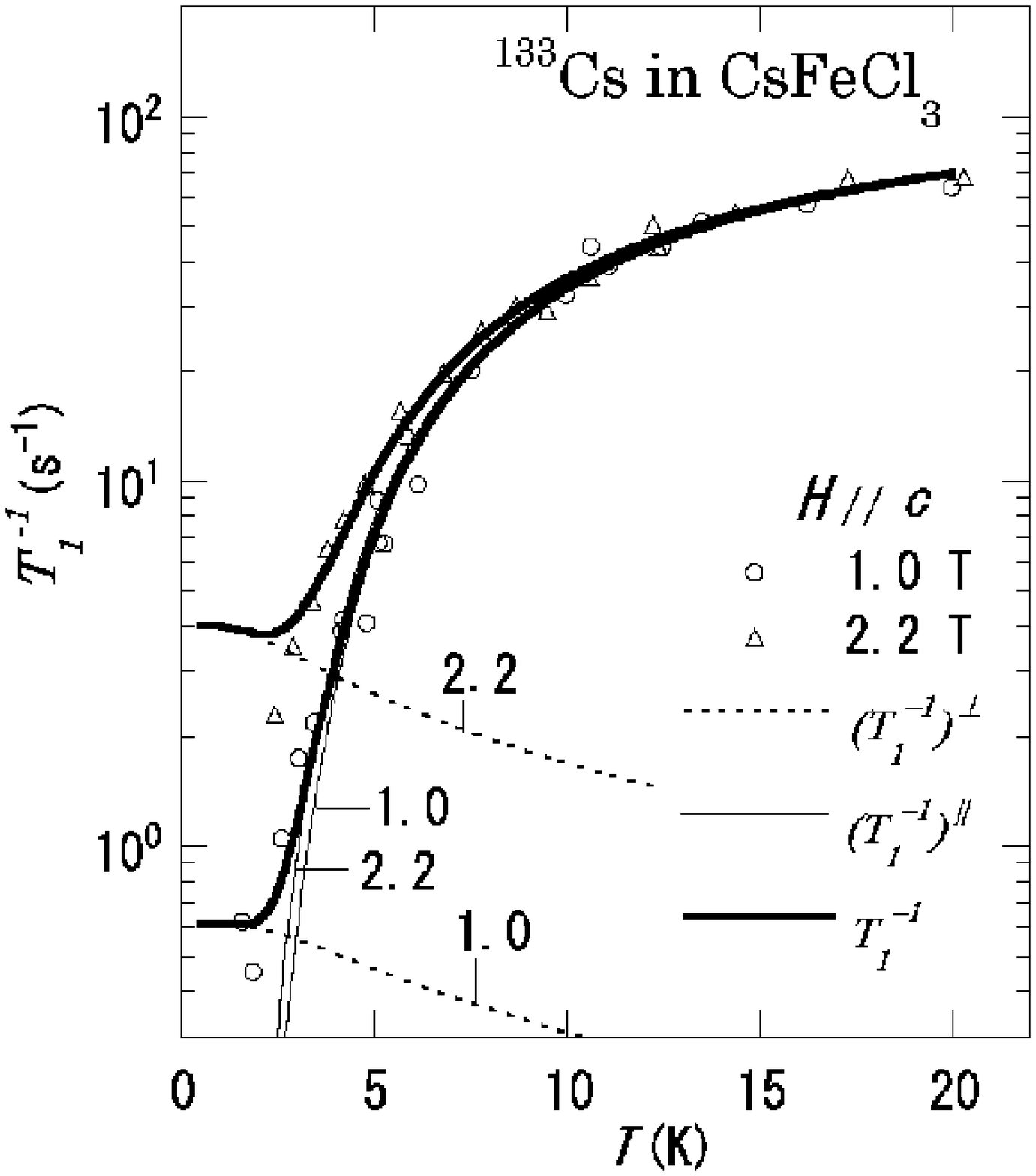}
  \end{center}
\caption{The results of fitting at $H$=1.0 T and 2.2 T for CFC. The solid lines and the dotted lines represent the fitted curves obtained from eqs.(\ref{Tz}) and (\ref{Txy}) in the text. The heavy lines represent the total relaxation rate $T_1^{-1} = (T_1^{-1})^{\mbox{\tiny $/\!\!/$}}+(T_1^{-1})^{\mbox{\tiny $\perp$}}$.}
\label{lowfield}
\end{figure}
 We show typical example of the ${\mib q}$ dependence of the static susceptibility $\chi^{\mbox{\tiny $\gamma$}}({\mib q})$ in Fig.\ref{kaixxparaq}, calculated at $H$=5.0 T. Figure \ref{kaixxparaq}(a) represents typical example of the ${\mib q}$ dependence of $\chi^{\mbox{\tiny $/\!\!/$}}({\mib q})$ along ${\mib q}=(1/3,1/3,q_c)$, each components being along the reciprocal-lattice vectors ${\mib a}^{\ast}$, ${\mib b}^{\ast}$ and ${\mib c}^{\ast}$, respectively. As is seen, $\chi^{\mbox{\tiny $/\!\!/$}}({\mib q})$ decreases appreciably with decreasing temperature, which results from the behavior of the longitudinal single-ion susceptibility $\phi^{\mbox{\tiny $/\!\!/$}}$($\omega$ = 0) characterizing the singlet ground state system. Figure \ref{kaixxparaq}(b) represents typical example of the ${\mib q}$ dependence of $\chi^{\mbox{\tiny $\perp$}}({\mib q})$ along ${\mib q}=(q,q,0)$. As is shown, $\chi^{\mbox{\tiny $\perp$}}({\mib q})$ exhibits a distinguished tendency of divergence around the ${\it K}$-point with decreasing temperature, which is associated with the softening of magnetic excitation $\omega_{\mbox{\footnotesize {\boldmath $q$}}}^{\mbox{\tiny $(-)$}}$.

 As an example, we show the ${\mib q}$-dependence of ${\cal A}^{\mbox{\tiny $/\!\!/$}}({\mib q})$ along ${\mib q}=(1/3,1/3,q_c)$ and ${\mib q}=(1/2,1/2,q_c)$ in Fig.\ref{Geoparaq}(a), and ${\cal A}^{\mbox{\tiny $\perp$}}({\mib q})$ along ${\mib q}=(q,q,0)$ in Fig. \ref{Geoparaq}(b). It should be noted that ${\cal A}^{\mbox{\tiny $/\!\!/$}}({\mib q})$ is zero and ${\cal A}^{\mbox{\tiny $\perp$}}({\mib q})$ takes the maximum at the {\it K}-point, which corresponds to the 120$^{\circ}$ spin-structure in the $c$-plane of the ordered state.
 
 In the following, we analyze the experimental results of $T_1^{-1}$ in CFC in terms of the contribution of $(T_1^{-1})^{\mbox{\tiny $/\!\!/$}}$ and $(T_1^{-1})^{\mbox{\tiny $\perp$}}$ (eqs.(\ref{Tz}) and (\ref{Txy})), with two variable parameters ${\it \Gamma_{\mbox{\footnotesize {\boldmath $q$}}}^{\mbox{\tiny $/\!\!/$}}}$ and ${\it \Gamma_{\mbox{\footnotesize {\boldmath $q$}}}^{\mbox{\tiny $\perp$}}}$. Here we assume that ${\it \Gamma_{\mbox{\footnotesize {\boldmath $q$}}}^{\mbox{\tiny $/\!\!/$}}}$ is independent of ${\mib q}$ and temperature, then the symbol ${\it \Gamma_{\mbox{\footnotesize {\boldmath $q$}}}^{\mbox{\tiny $/\!\!/$}}}$ is expressed as ${\it \Gamma^{\mbox{\tiny $/\!\!/$}}}$. As for ${\it \Gamma_{\mbox{\footnotesize {\boldmath $q$}}}^{\mbox{\tiny $\perp$}}}$, we assume that ${\it \Gamma_{\mbox{\footnotesize {\boldmath $q$}}}^{\mbox{\tiny $\perp$}}}$ is independent of temperature, and considering that $\chi^{\mbox{\tiny $\perp$}}({\mib K})$ increases largely and exhibits divergent behavior at low temperatures at the ${\it K}$-point, we put ${\it \Gamma_{\mbox{\footnotesize {\boldmath $q$}}}^{\mbox{\tiny $\perp$}}}$ $\approx$ ${\it \Gamma_{\mbox{\scriptsize ${\mib K}$}}^{\mbox{\tiny $\perp$}}}$.
\begin{figure}[b]
 \vspace{10pt}
  \begin{center}
    \epsfxsize=8.0cm    %  To keep the "height/width" ratio of your own figure,
    \epsfysize=9.0cm     %  please use either "epsfxsize" or "epsfysize" command.
    \epsfbox{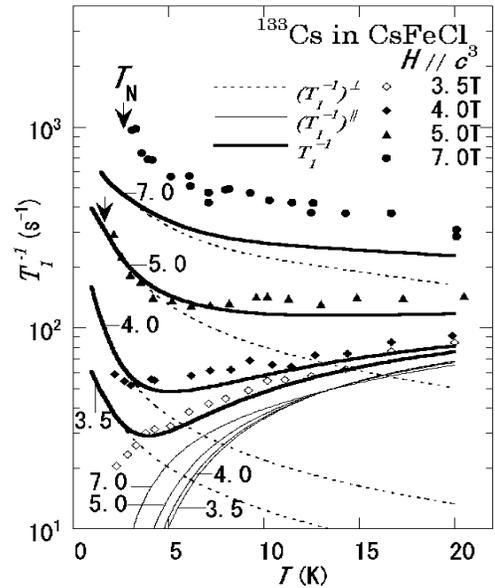}
  \end{center}
\caption{The results of fitting at $H$=3.5, 4.0, 5.0 and 7.0 T for CFC. The solid lines and the dotted lines represent the fitted curves obtained from eqs.(\ref{Tz}) and (\ref{Txy}) in the text. The heavy lines represent the total relaxation rate $T_1^{-1} = (T_1^{-1})^{\mbox{\tiny $/\!\!/$}}+(T_1^{-1})^{\mbox{\tiny $\perp$}}$.}
\label{highfield}
\end{figure}

 For the comparison, we show the analysis for the results in the low field region and high field region, respectively. First we look at the low field data which are shown in Fig.\ref{lowfield}. The heavy lines in Fig.\ref{lowfield} represent the best fitted curve of the total relaxation rate $T_1^{-1} = (T_1^{-1})^{\mbox{\tiny $/\!\!/$}}+(T_1^{-1})^{\mbox{\tiny $\perp$}}$. The solid lines represent the fitted curves of $(T_1^{-1})^{\mbox{\tiny $/\!\!/$}}$ obtained by choosing a constant value of $\hbar{\it \Gamma^{\mbox{\tiny $/\!\!/$}}}/k_{\mbox{\tiny B}}$ = 1.3 K, so as to fit the experimental results at low fields.
 Hereafter we leave out the factor $\hbar/k_{\mbox{\tiny B}}$. As is seen, the agreement between the data and the fitted curves of $(T_1^{-1})^{\mbox{\tiny $/\!\!/$}}$ is satisfactory over the whole temperature range at 1.0 T. Next, we look at the high field results which are shown in Fig.\ref{highfield}. The solid lines represent the calculated curves of $(T_1^{-1})^{\mbox{\tiny $/\!\!/$}}$ which are obtained by using the same value of ${\it \Gamma^{\mbox{\tiny $/\!\!/$}}}$ = 1.3 K. As is seen, the discrepancy between the experimental results and the calculation is serious. It is difficult to explain such a discrepancy by adjusting the value of ${\it \Gamma^{\mbox{\tiny $/\!\!/$}}}$, because the diverging behavior in $T_1^{-1}$ at low temperatures cannot be obtained from the calculatioon of $(T_1^{-1})^{\mbox{\tiny $/\!\!/$}}$. Such a discrepancy may be associated with another term of $(T_1^{-1})^{\mbox{\tiny $\perp$}}$. The dotted lines in Figs.\ref{lowfield} and \ref{highfield} represent the theoretical curves for $(T_1^{-1})^{\mbox{\tiny $\perp$}}$, which are obtained by choosing the values of ${\it \Gamma_{\mbox{\scriptsize ${\mib K}$}}^{\mbox{\tiny $\perp$}}}$ so as to fit the experimental results at low temperatures. The heavy lines in Fig.\ref{highfield} represent the total relaxation rate $T_1^{-1} = (T_1^{-1})^{\mbox{\tiny $/\!\!/$}}+(T_1^{-1})^{\mbox{\tiny $\perp$}}$. Then we find that the reasonable fit is attained by taking that ${\it \Gamma_{\mbox{\scriptsize ${\mib K}$}}^{\mbox{\tiny $\perp$}}}$ increases with increasing an external field, while a constant value is expected for ${\it \Gamma^{\mbox{\tiny $/\!\!/$}}}$. The best fit values of ${\it \Gamma_{\mbox{\scriptsize ${\mib K}$}}^{\mbox{\tiny $\perp$}}}$ at each field are summarized as follows; ${\it \Gamma_{\mbox{\scriptsize ${\mib K}$}}^{\mbox{\tiny $\perp$}}}$ $\approx$ 0.02, 0.08, 0.4, 0.6, 1.6 and 2.0 K (for $H$= 1.0, 2.2, 3.5, 4.0, 5.0 and 7.0 T). It turns out that there is a significant difference in the value of ${\it \Gamma_{\mbox{\scriptsize ${\mib K}$}}^{\mbox{\tiny $\perp$}}}$, which is very small in the low field region, and relatively large in the high field region. This fact suggests that the damping of magnetic excitation becomes remarkable in the high field region where the effect of the softening is more pronounced as the sign of the occurrence of 3D-LRO.
\begin{figure}[b]
 \vspace{10pt}
  \begin{center}
    \epsfxsize=8.0cm    %  To keep the "height/width" ratio of your own figure,
    \epsfysize=7.5cm     %  please use either "epsfxsize" or "epsfysize" command.
    \epsfbox{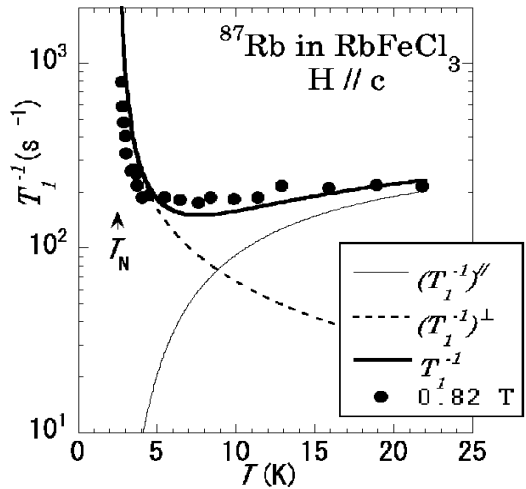}
  \end{center}
\caption{The fitted curves of $T_1^{-1}$ at $H$=0.82 T for RFC. The solid line and the dotted line represent the fitted curves obtained from eqs.(\ref{Tz}) and (\ref{Txy}) in the text. The heavy line represents the total relaxation rate $T_1^{-1} = (T_1^{-1})^{\mbox{\tiny $/\!\!/$}}+(T_1^{-1})^{\mbox{\tiny $\perp$}}$.}
\label{Rbpara82}
\end{figure}

 Next we consider the results of $T_1^{-1}$ in RFC. The analysis was performed following the same procedure in CFC. The value of ${\it \Gamma^{\mbox{\tiny $/\!\!/$}}}$ is determined so that the contribution of $(T_1^{-1})^{\mbox{\tiny $/\!\!/$}}$ explains the data at higher temperatures. The contribution of $(T_1^{-1})^{\mbox{\tiny $\perp$}}$ is determined so as to fit the curve of the total relaxation rate to the experimental results around $T_{\rm N}$. At $H$=0.82 T, the best fitted curves are obtained for ${\it \Gamma^{\mbox{\tiny $/\!\!/$}}}=$3.0 K and ${\it \Gamma_{\mbox{\scriptsize ${\mib K}$}}^{\mbox{\tiny $\perp$}}}=$0.4 K. As shown in Fig.\ref{Rbpara82}, the agreement between the theoretical curves and the experimental results is very good.
 
\subsection{$H \bot c$}
 In the case of $H \bot c$, we first consider the results in RFC. It is noted that the effect of the external field is more evident than in the case of CFC. 
\begin{figure}[h]  
 \begin{center}
    \epsfxsize=8.0cm    %  To keep the "height/width" ratio of your own figure,
    \epsfysize=6.2cm     %  please use either "epsfxsize" or "epsfysize" command.
    \epsfbox{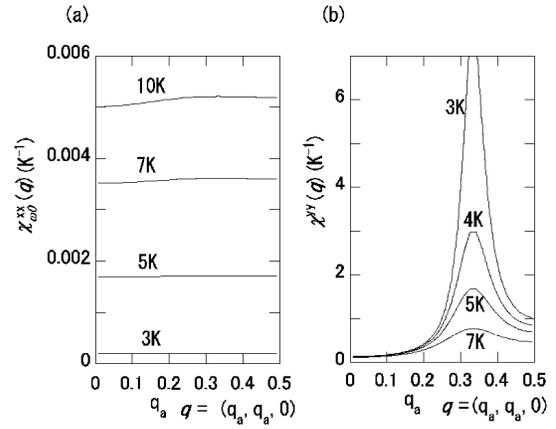}
  \end{center}
\caption{(a) Temperature dependences of $\chi^{xx}_{\omega_0}({\mib q})$ along ${\mib q}=(q,q,0)$ and (b) $\chi^{yy}({\mib q})$ along ${\mib q}=(q,q,0)$, calculated for RFC at $H$=1.0 T on the basis of the DCEFA.}
\label{kaixxkaiyy}
\end{figure}
\begin{figure}[b]  
 \begin{center}
    \epsfxsize=8.0cm    %  To keep the "height/width" ratio of your own figure,
    \epsfysize=6.2cm     %  please use either "epsfxsize" or "epsfysize" command.
    \epsfbox{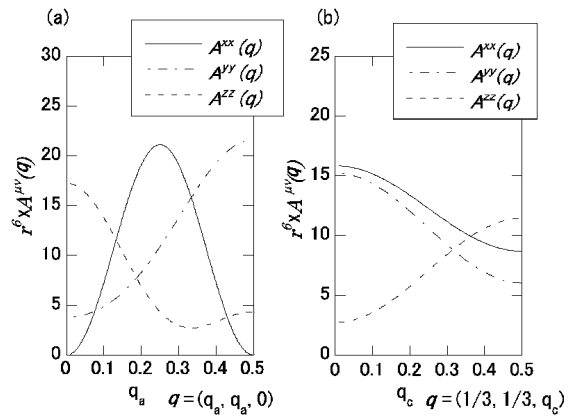}
  \end{center}  
\caption{(a) Geometrical factors $r^6 \times {\cal A}^{\mu\nu}({\mib q})$ along ${\mib q}=(q,q,0)$ and (b) $r^6 \times {\cal A}^{\mu\nu}({\mib q})$ along ${\mib q}=(1/3,1/3,q_c)$, where r is the distance between the $^{87}$Rb nucleus and the nearest Fe$^{2+}$ ions ($r=4.3\times10^{-8}$cm).}
\label{qp}
\end{figure}
 We show the typical example of the ${\mib q}$ dependence of $\chi_{\omega_0}^{xx}({\mib q})$ and $\chi^{yy}({\mib q})$ along ${\mib q}=(q,q,0)$ in Fig.\ref{kaixxkaiyy}, calculated at $H$=1.0 T. As is seen, $\chi^{xx}_{\omega_0}({\mib q})$ decreases appreciably with decreasing temperature, which mainly results from the behavior of $\phi^{xx}(\omega=0)$. The temperature dependence of $(T_1^{-1})_{\mbox{\tiny $0$}}$ follows the behavior of $\chi^{xx}_{\omega_0}({\mib q})T$, and decreases with decreasing temperature. On the other hand, as seen in Fig.\ref{kaixxkaiyy}(b), $\chi^{yy}({\mib q})$ increases with decreasing temperature, and exhibits a distinguished tendency of divergence around the {\it K}-point; this behavior is associated with the softening of magnetic excitation $\omega_{\mbox{\scriptsize ${\mib K}$}}^{\mbox{\tiny $(1)$}}$ and mainly comes from the contribution of $\chi^{yy}_{\omega_1}({\mib K})$.

 Figure \ref{qp} shows the calculated results of ${\cal A}^{xx}({\mib q})$, ${\cal A}^{yy}({\mib q})$ and ${\cal A}^{zz}({\mib q})$ along ${\mib q}=(q,q,0)$ (Fig.\ref{qp}(a)) and ${\mib q}=(1/3,1/3,q_c)$ (Fig.\ref{qp}(b)). It is noted that ${\cal A}^{yy}({\mbox{\footnotesize {\boldmath $K$}}})$ has large value around the $K$-point; therefore, the effect of the divergence of $\chi^{yy}_{\omega_1}({\mib K})$ will largely appear on $(T_1^{-1})_{\mbox{\tiny $1$}}$.
 
 In the following, we analyze the experimental results of $T_1^{-1}$ in terms of $(T_1^{-1})_{\mbox{\tiny $1$}}$, $(T_1^{-1})_{\mbox{\tiny $2$}}$ and $(T_1^{-1})_{\mbox{\tiny $0$}}$ (eqs.(\ref{T11}), (\ref{T12}) and (\ref{T0})), with ${\it \Gamma_{\mbox{\footnotesize {\boldmath $q$}}}^{\mbox{\tiny $(0)$}}}$, ${\it \Gamma_{\mbox{\footnotesize {\boldmath $q$}}}^{\mbox{\tiny $(1)$}}}$ and ${\it \Gamma_{\mbox{\footnotesize {\boldmath $q$}}}^{\mbox{\tiny $(2)$}}}$ as an adjustable parameters. We here assume that ${\it \Gamma_{\mbox{\footnotesize {\boldmath $q$}}}^{\mbox{\tiny $(0)$}}}$ and ${\it \Gamma_{\mbox{\footnotesize {\boldmath $q$}}}^{\mbox{\tiny $(2)$}}}$ are independent of ${\mib q}$ and temperature, thus the symbols ${\it \Gamma_{\mbox{\footnotesize {\boldmath $q$}}}^{\mbox{\tiny $(0)$}}}$ and ${\it \Gamma_{\mbox{\footnotesize {\boldmath $q$}}}^{\mbox{\tiny $(2)$}}}$ are expressed as ${\it \Gamma^{\mbox{\tiny $(0)$}}}$ and ${\it \Gamma^{\mbox{\tiny $(2)$}}}$. As for ${\it \Gamma_{\mbox{\footnotesize {\boldmath $q$}}}^{\mbox{\tiny $(1)$}}}$, we assume that ${\it \Gamma_{\mbox{\footnotesize {\boldmath $q$}}}^{\mbox{\tiny $(1)$}}}$ is independent of temperature, and considering that $\chi^{yy}_{\omega_1}({\mib K})$ increases largely and exhibits divergent behavior at low temperatures at the $K$-point, we put ${\it \Gamma_{\mbox{\footnotesize {\boldmath $q$}}}^{\mbox{\tiny $(1)$}}}\approx{\it \Gamma_{\mbox{\scriptsize ${\mib K}$}}^{\mbox{\tiny $(1)$}}}$. 
 
\begin{figure}[h]
 \vspace{10pt}
  \begin{center}
    \epsfxsize=8.5cm    %  To keep the "height/width" ratio of your own figure,
    \epsfysize=6.5cm     %  please use either "epsfxsize" or "epsfysize" command.
    \epsfbox{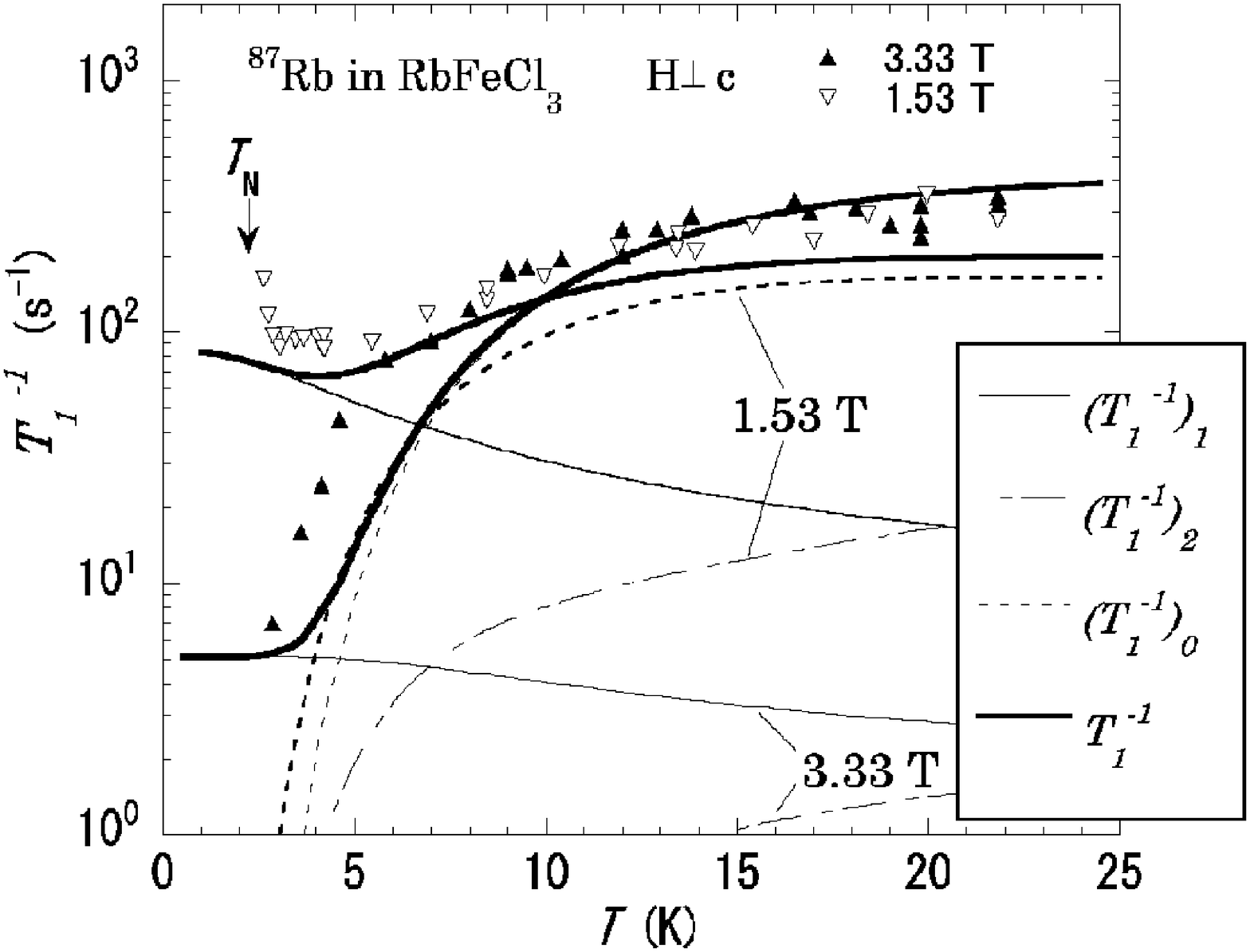}
  \end{center}
\caption{The results of fitting at $H$=1.53 T and 3.33 T ($H \bot c$) for RFC. The solid-lines, the dashed-dotted lines and the dotted lines represent the fitted curves obtained from eqs.(\ref{T11}), (\ref{T12}) and (\ref{T0}) in the text, respectively. The heavy lines represent the total relaxation rate $T_1^{-1}=(T_1^{-1})_{\mbox{\tiny $1$}} + (T_1^{-1})_{\mbox{\tiny $2$}} + (T_1^{-1})_{\mbox{\tiny $0$}}$.}
\label{Rbpe3T}
\end{figure}

 First, we assume that the contribution of $(T_1^{-1})_{\mbox{\tiny $2$}}$ is very small, because this is due to the excitation between the excited states.  Figure \ref{Rbpe3T} shows the fitted curves for the experimental results at 1.53 T and 3.33 T.
  When the external field of 3.33 T is applied in the $c$-plane, the excitation energy increases largely, and we may ignore the contribution of $(T_1^{-1})_{\mbox{\tiny $2$}}$. The solid line and the dotted line represent the fitted curves of $(T_1^{-1})_{\mbox{\tiny $1$}}$ and $(T_1^{-1})_{\mbox{\tiny $0$}}$, respectively. The value of $(T_1^{-1})_{\mbox{\tiny $0$}}$ decreases, while that of $(T_1^{-1})_{\mbox{\tiny $1$}}$ increases with decreasing temperature; therefore, at 3.33 T, the analysis was performed assuming that the relaxation rate is mainly due to the contribution of $(T_1^{-1})_{\mbox{\tiny $0$}}$ at higher temperatures, and the descrepancy from the data at low temperatures is ascribed to the contribution of $(T_1^{-1})_{\mbox{\tiny $1$}}$. The best fit for the data at 3.33 T is obtained with ${\it \Gamma^{\mbox{\tiny $(0)$}}}=5.0$ K and ${\it \Gamma^{\mbox{\tiny $(1)$}}_{\mbox{\scriptsize ${\mib K}$}}}=0.4$ K. The heavy line represents the total relaxation rate $T_1^{-1}=(T_1^{-1})_{\mbox{\tiny $1$}}+(T_1^{-1})_{\mbox{\tiny $2$}}+(T_1^{-1})_{\mbox{\tiny $0$}}$, and the estimation of the small contribution of $(T_1^{-1})_{\mbox{\tiny $2$}}$ will be explained later. At 1.53 T, the phase transition occurs, and the tendency of divergence of $T_1^{-1}$ appears at around $T_{\rm N}$. We expect that this behavior results from the contribution of $(T_1^{-1})_{\mbox{\tiny $1$}}$. Refering to the fitting procedure of ${\it \Gamma_{\mbox{\scriptsize ${\mib K}$}}^{\mbox{\tiny $\perp$}}}$ in CFC, we expect that the damping constant of the soft mode (${\it \Gamma^{\mbox{\tiny $(1)$}}_{\mbox{\scriptsize ${\mib K}$}}}$) depends on the field and the occurrence of the magnetic phase transition. The best fit is obtained when the value of ${\it \Gamma^{\mbox{\tiny $(1)$}}_{\mbox{\scriptsize ${\mib K}$}}}$ is varied from 0.4 K to 1.5 K, and we take it that damping of $\omega^{\mbox{\tiny $(1)$}}_{\mbox{\scriptsize ${\mib K}$}}$ mode becomes severe as the sign of the occurrence of 3D-LRO.

\begin{figure}[b]
 \vspace{10pt}
  \begin{center}
    \epsfxsize=8.5cm    %  To keep the "height/width" ratio of your own figure,
    \epsfysize=6.5cm     %  please use either "epsfxsize" or "epsfysize" command.
    \epsfbox{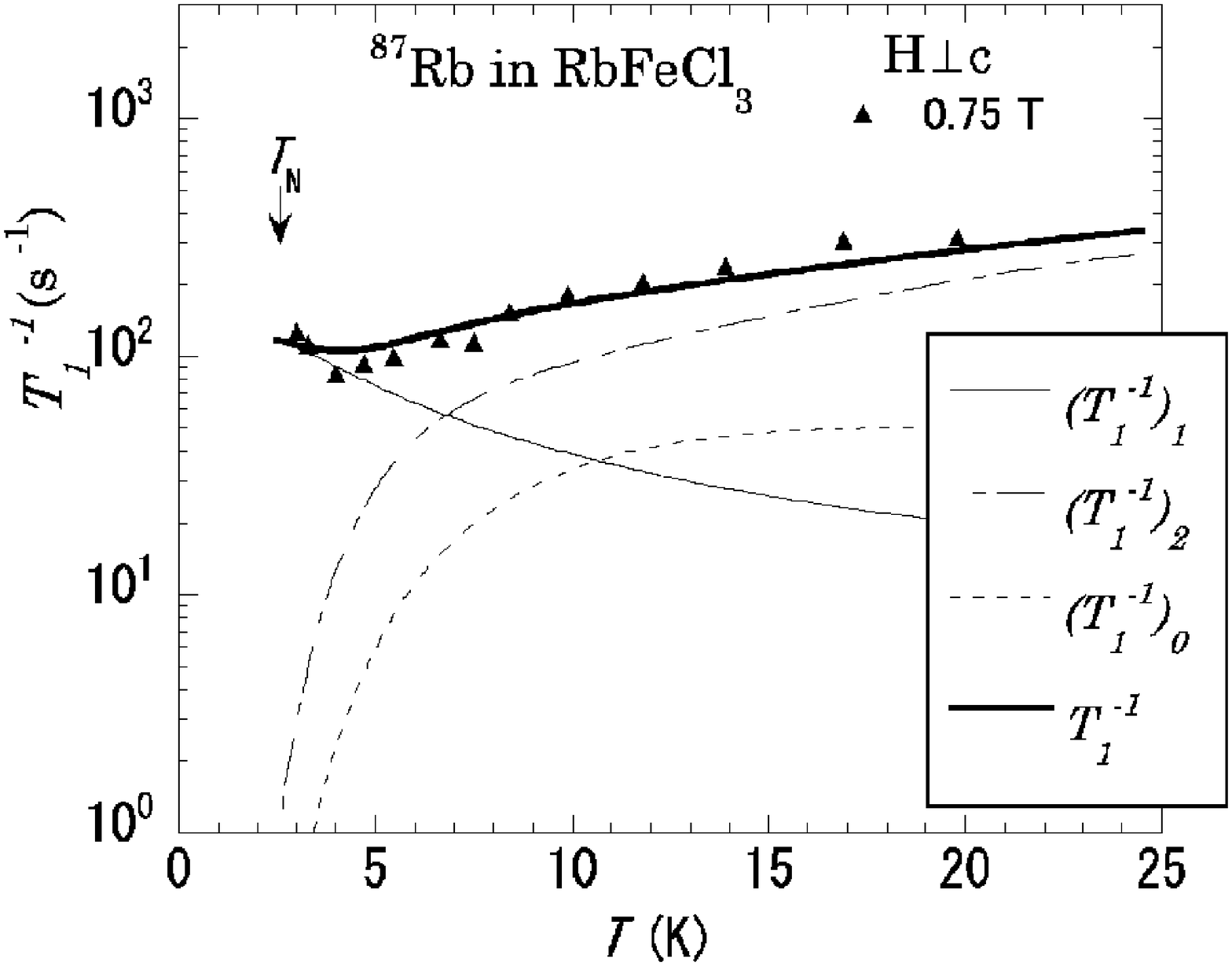}
  \end{center}
\caption{The results of fitting at $H$=0.75 T ($H \bot c$) for RFC. The solid-line, the dashed-dotted line and the dotted line represent the fitted curves obtained from eqs.(\ref{T11}), (\ref{T12}) and (\ref{T0}) in the text, respectively. The heavy line represents the total relaxation rate $T_1^{-1}=(T_1^{-1})_{\mbox{\tiny $1$}} + (T_1^{-1})_{\mbox{\tiny $2$}} + (T_1^{-1})_{\mbox{\tiny $0$}}$.}
\label{Rbperp75}
\end{figure}
 Now we turn to the data at 0.75 T, and fit the theoretical curves using the values of ${\it \Gamma^{\mbox{\tiny $(0)$}}}$ and ${\it \Gamma^{\mbox{\tiny $(1)$}}_{\mbox{\scriptsize ${\mib K}$}}}$ obtained at 1.53 T. However, it turns out that the reasonable fit cannot be obtained at 0.75 T within the contributions of $(T_1^{-1})_{\mbox{\tiny $1$}}$ and $(T_1^{-1})_{\mbox{\tiny $0$}}$, especially at higher temperatures. Considering that $\omega^{\mbox{\tiny $(2)$}}_{\mbox{\footnotesize ${\mib q}$}}$ is the excitation between the excited states, and $(T_1^{-1})_{\mbox{\tiny $2$}}$ increases with increasing temperature, now we take into account of the contribution of $(T_1^{-1})_{\mbox{\tiny $2$}}$. 

 Figure \ref{Rbperp75} shows the fitted curves for the experimental results at 0.75 T. The dashed-dotted line represent the fitted curve of $(T_1^{-1})_{\mbox{\tiny $2$}}$, and the best fit is obtained when the contribution of $(T_1^{-1})_{\mbox{\tiny $2$}}$ is included with ${\it \Gamma^{\mbox{\tiny $(2)$}}}=0.04$ K. It is conjectured that magnetic excitation of $\omega^{\mbox{\tiny $(2)$}}_{\mbox{\footnotesize ${\mib q}$}}$ mode, which do not exist in case of $H {\mbox {\scriptsize $/\!\!/$}} c$, would have appreciable effect on $T_1^{-1}$. In Fig.\ref{Rbpe3T}, we have also shown the fitted curves of $(T_1^{-1})_{\mbox{\tiny $2$}}$ and the total relaxation rate at 1.53 T and 3.33 T with ${\it \Gamma^{\mbox{\tiny $(2)$}}}= 0.04$ K. 
\begin{figure}[b]
 \vspace{10pt}
  \begin{center}
    \epsfxsize=8.5cm    %  To keep the "height/width" ratio of your own figure,
    \epsfysize=7.0cm     %  please use either "epsfxsize" or "epsfysize" command.
    \epsfbox{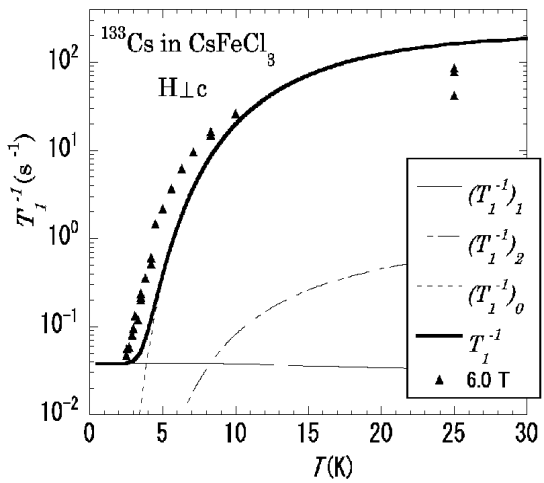}
  \end{center}
\caption{The results of fitting at $H$=6.0 T ($H \bot c$) for CFC. The solid-line, the dashed-dotted line and the dotted line represent the fitted curves obtained from eqs.(\ref{T11}), (\ref{T12}) and (\ref{T0}) in the text, respectively. The heavy line represents the total relaxation rate $T_1^{-1}=(T_1^{-1})_{\mbox{\tiny $1$}} + (T_1^{-1})_{\mbox{\tiny $2$}} + (T_1^{-1})_{\mbox{\tiny $0$}}$.}
\label{Csperp6T}
\end{figure} 

 Next we analyze the experimental results of $T_1^{-1}$ of $^{133}$Cs in CFC. Figure \ref{Csperp6T} shows the fitted curves of $T_1^{-1}$ at 6.0 T. Referring to the analysis for the data at 3.33 T in RFC, the analysis was performed comparing the contributions of $(T_1^{-1})_{\mbox{\tiny $0$}}$ and $(T_1^{-1})_{\mbox{\tiny $1$}}$, because the contribution of $(T_1^{-1})_{\mbox{\tiny $2$}}$ can be still more disregarded owing to the larger effect of the external field. The value of ${\it \Gamma^{\mbox{\tiny $(0)$}}}$ is determined so that the contribution of $(T_1^{-1})_{\mbox{\tiny $0$}}$ explains the data at higher temperatures, and the value of ${\it \Gamma^{\mbox{\tiny $(1)$}}_{\mbox{\scriptsize ${\mib K}$}}}$ is determined so as to fit the curve of the total relaxation rate to the data around 2 K. The value of ${\it \Gamma^{\mbox{\tiny $(2)$}}}$ follows that obtained in RFC, because this value will not be largely different between these materials. The best fitted curves are obtained with ${\it \Gamma^{\mbox{\tiny $(1)$}}_{\mbox{\scriptsize ${\mib K}$}}}=0.07$ K, ${\it \Gamma^{\mbox{\tiny $(2)$}}}=0.04$ K and ${\it \Gamma^{\mbox{\tiny $(0)$}}}=2.5$ K, respectively.

\subsection{Discussion} 
 Throughout the analyses based on the DCEFA, we have evaluated quantitatively the damping constants of the spin fluctuations which have effects on the relaxation mechanism. The damping constants for the fluctuation of the induced moment are determined at somewhat smaller values than those of the relative exchange interactions. The lifetime of magnetic excitation is long and the values of damping constants are evaluated more or less about 0.1 K, in the field region where the system remains non-magnetic down to zero K. On the other hand, in the field region where the ordering occurs, it was deduced that the lifetime of soft mode is short, and the values of damping constants are estimated as large as 1.5 $\sim$ 2.0 K. If we consider that there is interaction between magnetic excitations, then the collisions between excitations will explain the field dependence of ${\it \Gamma_{\mbox{\scriptsize ${\mib K}$}}^{\mbox{\tiny $\perp$}}}$ (${\it \Gamma_{\mbox{\scriptsize ${\mib K}$}}^{\mbox{\tiny $(1)$}}}$). When the magnetic phase transition occurs at low temperatures, then the times of collisions might increase at $T \sim T_{\rm N}$; because large numbers of the excitations overcome the activation energy when the excitation energy decreases to zero. Then owing to the effect of the anharmonic terms of the interactions, lifetime of magnetic excitation will become short and the value of damping constant will increase. 

 The evaluated values of damping constants are summarized in Table \ref{damp val}. It is very instructive to refer to the experimental results of inelastic neutron scattering on CFC, which have measured the line width of the excitation spectra under various external fields applied parallel to the $c$-axis at $T$=1.6 K\cite{rf:30}. The line widths of $\omega^{\mbox{\tiny $(-)$}}_{\mbox{\scriptsize ${\mib K}$}}$ mode increases largely in the field region where the magnetic phase transition occurs, and the absolute value of the line width is about 1 K above 4.5 T. This value is only different from the estimated value of ${\it \Gamma_{\mbox{\scriptsize ${\mib K}$}}^{\mbox{\tiny $\perp$}}}$ for the factor of 2 at the maximum. 
\begin{table}[b]
 \caption{The values of damping constants evaluated through the fit.}% {}内に表題を書く
 \begin{center}
  \begin{tabular}{@{\hspace{\tabcolsep}\extracolsep{\fill}}ccc}
    \hline
       &CsFeCl$_3$ & RbFeCl$_3$  \\
    \hline
     ${\it \Gamma_{\mbox{\tiny ${\mib K}$}}^{\mbox{\tiny $\perp$}}}$ (K)  &  0.02$\sim$2.0  &  0.4  \\
     ${\it \Gamma^{\mbox{\tiny $/\!\!/$}}}$ (K) &  1.3  &  3.0  \\
     ${\it \Gamma^{\mbox{\tiny $(1)$}}_{\mbox{\tiny ${\mib K}$}}}$ (K) &  0.07  &  0.4$\sim$1.5  \\
     ${\it \Gamma^{\mbox{\tiny $(2)$}}}$ (K) &  0.04  &  0.04  \\
     ${\it \Gamma^{\mbox{\tiny $(0)$}}}$ (K) &  2.5  & 5.0   \\
    \hline
  \end{tabular}
 \end{center}
 \label{damp val}
\end{table}
 
 We emphasize that various values of damping constants shown in Table \ref{damp val} are new informations, which are obtained for the first time through the analyses of $T_1^{-1}$ in the singlet ground state system. In particular, it is noteworthy that we could estimate the damping of the longitudinal fluctuation and the excitation of $\omega^{\mbox{\tiny $(2)$}}_{\mbox{\footnotesize ${\mib q}$}}$ modes, and show their effects on $T_1^{-1}$; inelastic neutron scattering technique is difficult for investigating the fluctuation at $\omega \approx 0$, and the excitation between the excited states. Informations of such fluctuations are obtained through the advantages of NMR experiment. 

\section{Conclusion}
 In conclusion, we have measured the temperature dependences of $T_1^{-1}$ of $^{133}$Cs ($^{87}$Rb) in the singlet ground state system CFC(RFC) under various fields ($H$$\leq$7 T) applied parallel or perpendicular to the $c$-axis. In the field region where the system remains non-magnetic down to zero K, it was found that a remarkable decrease of $T_1^{-1}$ was governed by the longitudinal fluctuation along the external field, which depends strongly on the appreciable energy gap between the ground state and the excited states. On the contrary, in the field region where a magnetic ordering occurs, an appreciable increase of $T_1^{-1}$ below about 5 K is ascribed to the transverse fluctuation perpendicular to the external field, which is associated with the softening of magnetic excitation. The damping constants of the normalized relaxation function were evaluated for the fluctuations which have significant effects on the nuclear magnetic relaxation. It was deduced that the damping constant of the soft mode increases largely in the field region where the softening occurs, and then the tendency of divergence of $T_1^{-1}$ is ascribed to that of the static susceptibility. In RFC, it was conjectured that the magnetic excitation of $\omega^{\mbox{\tiny $(2)$}}_{\mbox{\footnotesize ${\mib q}$}}$ mode would have significant effects on $T_1^{-1}$ in case of $H \bot c$. 

 We thank Mr. S. Ueda for cooperation with the preparation of sample.

\appendix
\section{Dynamical Susceptibitlies}
The dynamical single-ion susceptibilities are calculated by the following expressions;
\begin{equation}
\phi^{\mu\nu}(\omega)=\sum_{nm}\frac{(\rho_n-\rho_m)<m|S_{i\mu}|n><n|S_{i\nu}|m>}{\omega+E_m-E_n},
\label{singleionsus}
\end{equation}   
with $\rho_n=\exp(-E_n/k_{\mbox{\tiny B}}T)/\sum_m \exp(-E_m/k_{\mbox{\tiny B}}T)$ ({\it n},{\it m} = 1, 2 and 3). For the diagonal elements $n=m$, which exist for $S_{iz}$ in case of $H {\mbox{\footnotesize $/\!\!/$}} c$ and $S_{ix}$ in case of $H \bot c$, the dynamical single-ion susceptibilities are derived by considering the effect of the infinitesimal displacement of the effective field. Then for the case of $H {\mbox{\footnotesize $/\!\!/$}} c$, we obtain
\begin{equation}
\phi^{\mbox{\tiny $/\!\!/$}}(\omega) \equiv \phi^{zz}(\omega)=2(\frac{\rho_2+\rho_3}{k_{\mbox{\tiny B}}T}-\frac{\langle S_z \rangle^2}{k_{\mbox{\tiny B}} T})\delta_{\omega,0},
\label{phipara}
\end{equation}
and the dynamical susceptbilities $\chi^{\mbox{\tiny $/\!\!/$}}({\mib q},\omega)$ and $\chi^{\mbox{\tiny $\perp$}}({\mib q},\omega)$ for the whole system are expressed in terms of the single-ion susceptibilities as follows\cite{rf:20}:
\begin{equation}
\chi^{\gamma}({\mib q}, \omega)=\frac{\phi^{\gamma}(\omega)}{1-p^{\gamma}(J^{\gamma}({\mib q})-\alpha J^{\gamma}(0))\phi^{\gamma}(\omega)},
\label{kaiz}
\end{equation}
where $p^{\mbox{\tiny $/\!\!/$}}=2$ or $p^{\mbox{\tiny $\perp$}}=1$. 
 For example, the expression for $\chi^{\mbox{\tiny $\perp$}}({\mib q}, \omega)$ is rewritten in the following form;
\begin{equation}
\chi^{\mbox{\tiny $\perp$}}({\mib q}, \omega)=-\frac{2(\rho_{\mbox{\tiny $23$}}\omega+\rho_{\mbox{\tiny $12$}}E_{\mbox{\tiny $31$}} + \rho_{\mbox{\tiny $13$}}E_{\mbox{\tiny $21$}})}{(\omega+\omega_{\mbox{\footnotesize {\boldmath $q$}}}^{\mbox{\tiny $(-)$}})(\omega-\omega_{\mbox{\footnotesize {\boldmath $q$}}}^{\mbox{\tiny $(+)$}})},
\label{kaixq}
\end{equation}
where $\rho_{mn}=\rho_m-\rho_n, E_{mn}=E_m-E_n$ and $\omega_{\mbox{\footnotesize {\boldmath $q$}}}^{\mbox{\tiny $(\pm)$}}$ is expressed as
\begin{eqnarray}
\omega_{\mbox{\footnotesize {\boldmath $q$}}}^{\mbox{\tiny $(\pm)$}}& &=\bigl[E^2-2E(J^{\mbox{\tiny $\perp$}}({\mib q})-\alpha J^{\mbox{\tiny $\perp$}}(0))(2-3\langle S_z \rangle^2)\nonumber \\
+& &\langle S_z \rangle^2(J^{\mbox{\tiny $\perp$}}({\mib q})-\alpha J^{\mbox{\tiny $\perp$}}(0))^2]^{(1/2)}
 \pm[g^{\mbox{\tiny $/\!\!/$}}\mu_{\mbox{\tiny B}}H \nonumber \\
+& &2J^{\mbox{\tiny $/\!\!/$}}(0)(1-\alpha)\langle S_z \rangle - (J^{\mbox{\tiny $\perp$}}({\mib q})-\alpha J^{\mbox{\tiny $\perp$}}(0))\langle S_z \rangle \bigr].\nonumber \\
& &
\label{omega-}
\end{eqnarray}

 In the case of $H \bot c$, the matrix representation of $S_{ix}$ have both off-diagonal and diagonal elements, then the dynamical single-ion susceptibilities are respectively given as
\begin{eqnarray}
\label{phixx}
\phi^{xx}(\omega)&=&\left\{ \begin{array}{ll}
            -\frac{{\textstyle 2\rho_{\mbox{\tiny $13$}}E_{\mbox{\tiny $31$}}(E/W)^2}}{{\textstyle \omega^2-E_{\mbox{\tiny $31$}}^2}},\hspace{3mm} (\omega \neq 0) \hspace{20mm}\vspace{2mm} \\
            -\frac{{\textstyle 2\rho_{\mbox{\tiny $13$}}(E/W)^2}}{{\textstyle E_{\mbox{\tiny $31$}}}} \hspace{20mm} \\
                          \end{array}
                  \right. \\
     & &\hspace{3mm} +\frac{\textstyle 1}{\textstyle k_{\mbox{\tiny B}} T}[(2B/W)^2(\rho_{\mbox{\tiny $1$}}+\rho_{\mbox{\tiny $3$}})-\langle S_x \rangle^2], (\omega = 0). \nonumber                
\end{eqnarray}                 
$\phi^{yz}(\omega)$, $\phi^{yy}(\omega)$ and $\phi^{zz}(\omega)$ are derived by using eq.(\ref{Sxmatrix}) and eq.(\ref{singleionsus}), and other components of the single-ion susceptibility are vanishing. The dynamical susceptbilities $\chi^{\mu\nu}({\mib q},\omega)$ for the whole spin system are expressed in terms of the single-ion susceptibilities as follows:
\begin{equation}  
\chi^{xx}({\mib q},\omega)=\phi^{xx}(\omega)/(1-2J_{\alpha}^{\mbox{\tiny $\perp$}}({\mib q})\phi^{xx}(\omega)),                
\label{chixxqomega}
\end{equation}
\begin{eqnarray}
\chi^{yy}({\mib q},\omega)&=&[\phi^{yy}(\omega)(1-2J_{\alpha}^{\mbox{\tiny $/\!\!/$}}({\mib q})\phi^{zz}(\omega))\nonumber \\
&+&2J_{\alpha}^{\mbox{\tiny $/\!\!/$}}({\mib q})\phi^{yz}(\omega)\phi^{zy}(\omega)]/D({\mib q},\omega),
\end{eqnarray}
\begin{equation}
\chi^{yz}({\mib q},\omega)=\phi^{yz}(\omega)/D({\mib q},\omega)=-\chi^{zy}({\mib q},\omega),
\label{chiyz}
\end{equation}
\begin{eqnarray}
\chi^{zz}({\mib q},\omega)&=&[\phi^{zz}(\omega)(1-2J_{\alpha}^{\mbox{\tiny $\perp$}}({\mib q})\phi^{yy}(\omega))\nonumber \\
&+&2J_{\alpha}^{\mbox{\tiny $\perp$}}({\mib q})\phi^{yz}(\omega)\phi^{zy}(\omega)]/D({\mib q},\omega),
\end{eqnarray}
 where
\begin{eqnarray}  
J_{\alpha}^{\mbox{\tiny $\gamma$}}({\mib q})&=&J^{\mbox{\tiny $\gamma$}}({\mib q})-\alpha J^{\mbox{\tiny $\gamma$}}(0), \\
D({\mib q},\omega)&=&1-2J_{\alpha}^{\mbox{\tiny $\perp$}}({\mib q})\phi^{yy}(\omega)-2J_{\alpha}^{\mbox{\tiny $/\!\!/$}}({\mib q})\phi^{zz}(\omega)\nonumber \\
&+&4J_{\alpha}^{\mbox{\tiny $/\!\!/$}}({\mib q})J_{\alpha}^{\mbox{\tiny $\perp$}}({\mib q})(\phi^{yy}(\omega)\phi^{zz}(\omega)-\phi^{yz}(\omega)\phi^{zy}(\omega)).\nonumber \\
& &
\label{Dqomega}
\end{eqnarray}
For example, $\chi^{yy}({\mib q},\omega)$ can be expressed as follows, 
\begin{eqnarray}
\hspace{-5mm}\chi^{yy}({\mib q},\omega)&=&\chi_{\omega_1}^{yy}({\mib q},\omega)+\chi^{yy}_{\omega_2}({\mib q},\omega)\nonumber \\
&=&-\biggl(\frac{I^{yy}_1}{\omega^2-(\omega^{\mbox{\tiny $(1)$}}_{\mbox{\footnotesize {\boldmath $q$}}})^2}+\frac{I^{yy}_2}{\omega^2-(\omega^{\mbox{\tiny $(2)$}}_{\mbox{\footnotesize {\boldmath $q$}}})^2}\biggl),
\label{twopole}
\end{eqnarray}
where $I^{yy}_1$ and $I^{yy}_2$ are given as 
\begin{equation}
I^{yy}_1=\frac{a_{\mbox{\tiny $1$}}(\omega^{\mbox{\tiny $(1)$}}_{\mbox{\footnotesize {\boldmath $q$}}})^2-b_{\mbox{\tiny $1$}}}{(\omega^{\mbox{\tiny $(1)$}}_{\mbox{\footnotesize {\boldmath $q$}}})^2-(\omega^{\mbox{\tiny $(2)$}}_{\mbox{\footnotesize {\boldmath $q$}}})^2}, I^{yy}_2=\frac{a_{\mbox{\tiny $1$}}(\omega^{\mbox{\tiny $(2)$}}_{\mbox{\footnotesize {\boldmath $q$}}})^2-b_{\mbox{\tiny $1$}}}{(\omega^{\mbox{\tiny $(2)$}}_{\mbox{\footnotesize {\boldmath $q$}}})^2-(\omega^{\mbox{\tiny $(1)$}}_{\mbox{\footnotesize {\boldmath $q$}}})^2},
\end{equation}
and $a_{\mbox{\tiny $1$}}$, $b_{\mbox{\tiny $1$}}$ are given as 
\begin{eqnarray}
a_{\mbox{\tiny $1$}}&=&2(\rho_{\mbox{\tiny $12$}}d_{\mbox{\tiny $1$}}^2 E_{\mbox{\tiny $21$}}+\rho_{\mbox{\tiny $23$}}d_{\mbox{\tiny $2$}}^2 E_{\mbox{\tiny $32$}})-8\rho_{\mbox{\tiny $13$}}^2 c_{\mbox{\tiny $1$}}^2 d_{\mbox{\tiny $1$}}^2 2J^{\mbox{\tiny $/\!\!/$}}({\mib q}), \nonumber \\
 & & \\
b_{\mbox{\tiny $1$}}&=&E_{\mbox{\tiny $21$}}^2 E_{\mbox{\tiny $32$}}^2 2(\frac{\rho_{\mbox{\tiny $12$}}d_{\mbox{\tiny $1$}}^2}{E_{\mbox{\tiny $21$}}}+\frac{\rho_{\mbox{\tiny $23$}}d_{\mbox{\tiny $2$}}^2}{E_{\mbox{\tiny $32$}}})\nonumber \\
& &\hspace{10mm}\times[1-4(\frac{\rho_{\mbox{\tiny $12$}}c_{\mbox{\tiny $1$}}^2}{E_{\mbox{\tiny $21$}}}+\frac{\rho_{\mbox{\tiny $23$}}c_{\mbox{\tiny $2$}}^2}{E_{\mbox{\tiny $32$}}})2J^{\mbox{\tiny $/\!\!/$}}({\mib q})].
\end{eqnarray}
 $\omega^{\mbox{\tiny $(1)$}}_{\mbox{\footnotesize {\boldmath $q$}}}$ and $\omega^{\mbox{\tiny $(2)$}}_{\mbox{\footnotesize {\boldmath $q$}}}$ are given as 
\begin{equation}
\omega^{\mbox{\tiny $(1)$}}_{\mbox{\footnotesize {\boldmath $q$}}}=A_{\mbox{\tiny $1$}}+(A_{\mbox{\tiny $1$}}^2-B_{\mbox{\tiny $1$}})^{(1/2)}, \omega^{\mbox{\tiny $(2)$}}_{\mbox{\footnotesize {\boldmath $q$}}}=A_{\mbox{\tiny $1$}}-(A_{\mbox{\tiny $1$}}^2-B_{\mbox{\tiny $1$}})^{(1/2)},
\end{equation}
where $A_{\mbox{\tiny $1$}}$ and $B_{\mbox{\tiny $1$}}$ are given as 
\begin{eqnarray}
& &A_1=\frac{1}{2}(E_{\mbox{\tiny $21$}}^2+E_{\mbox{\tiny $32$}}^2)
   -(\rho_{\mbox{\tiny $12$}} E_{\mbox{\tiny $21$}} d_{\mbox{\tiny $1$}}^2 + \rho_{\mbox{\tiny $23$}} E_{\mbox{\tiny $32$}} d_{\mbox{\tiny $2$}}^2) 2J^{\mbox{\tiny $\perp$}}({\mib q}) \nonumber \\
   &-& (\rho_{\mbox{\tiny $12$}} E_{\mbox{\tiny $21$}} c_{\mbox{\tiny $1$}}^2 + \rho_{\mbox{\tiny $23$}} E_{\mbox{\tiny $32$}} c_{\mbox{\tiny $2$}}^2) 4J^{\mbox{\tiny $/\!\!/$}}({\mib q})
   + 16 \rho_{\mbox{\tiny $13$}}^2 c_{\mbox{\tiny $1$}}^2 d_{\mbox{\tiny $1$}}^2 J^{\mbox{\tiny $/\!\!/$}}({\mib q})
   J^{\mbox{\tiny $\perp$}}({\mib q}), \nonumber \\
& & \\
& &B_{\mbox{\tiny $1$}}=E_{\mbox{\tiny $21$}}^2E_{\mbox{\tiny $32$}}^2[1-2(\frac{\rho_{\mbox{\tiny $12$}}d_{\mbox{\tiny $1$}}^2}{E_{\mbox{\tiny $21$}}}+\frac{\rho_{\mbox{\tiny $23$}}d_{\mbox{\tiny $2$}}^2}{E_{\mbox{\tiny $32$}}})
2J^{\mbox{\tiny $\perp$}}({\mib q})] \nonumber \\
& &\hspace{20mm}\times[1-4(\frac{\rho_{\mbox{\tiny $12$}}c_{\mbox{\tiny $1$}}^2}{E_{\mbox{\tiny $21$}}}
+\frac{\rho_{\mbox{\tiny $23$}}c_{\mbox{\tiny $2$}}^2}{E_{\mbox{\tiny $32$}}})2J^{\mbox{\tiny $/\!\!/$}}({\mib q})].
\end{eqnarray}

\section{Geometrical Factors}
 The calculations for the geometrical factors were made by taking the dipolar sum only over the nearest-neighbor sites of the Fe$^{2+}$ ions, considering the fact that this factor decreases as inverse sixth power of the distance between the $^{133}$Cs ($^{87}$Rb) nucleus and the Fe$^{2+}$ ion. 

\begin{figure}[h]
 \vspace{10pt}
  \begin{center}
    \epsfxsize=7.0cm    %  To keep the "height/width" ratio of your own figure,
    \epsfysize=3.5cm     %  please use either "epsfxsize" or "epsfysize" command.
    \epsfbox{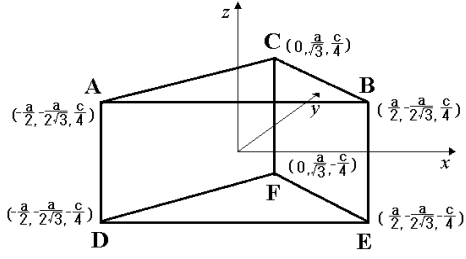}
  \end{center}
   \vspace{10pt}
\caption{The illustration of geometrical figure of the $^{133}$Cs ($^{87}$Rb) nucleus and the nearest neighbor Fe$^{2+}$ ions. The position of the $^{133}$Cs ($^{87}$Rb) nucleus is taken at the origin, then the sites of the Fe$^{2+}$ ions on the vertexes of the trigonal prism are labeled from A to F. The coordinates of each site are given using the  crystal parameters $a$ and $c$.}
\label{Trigonal}
\end{figure}

 For the convenient, we consider the geometrical figure as shown in Fig.\ref{Trigonal}. The position of $^{133}$Cs ($^{87}$Rb) is taken at the origin, and the sites of the Fe$^{2+}$ ions are placed on the vertexes of the trigonal prism, which are labeled from A to F. The coordinates of each site are given using the lattice parameters $a$ and $c$.
 The $\mu^{\prime}$-component of the dipole field (${\mib h}_i$) at the $^{133}$Cs ($^{87}$Rb) nucleus, due to the $\mu$-component of the magnetic moment (${\mib m}_i$) of Fe$^{2+}$ ion at the site $i$, is given as follows;
\begin{equation}
h^{\mu^{\prime}}_i=(\frac{\delta^{\mu^{\prime}\mu}}{r_i^3}-3\frac{r^{\mu^{\prime}}r^{\mu}}{r_i^5})m^{\mu}_i=d_{\mu^{\prime}\mu}(r_i)m^{\mu}_i,
\label{dipole}
\end{equation}
where $r_i$ is the distance from the origin to ${\mib m}_i$, and $d_{\mu^{\prime}\mu}(r_i)$ is the dipole tensor. Then the correlation of the fluctuating local field is related to the spin correlation via the product of $d_{\mu^{\prime}\mu}(r_i)$ as
\begin{equation}
\langle \delta h^{\mu^{\prime}}(t) \delta h^{\nu^{\prime}}(0)\rangle=\sum_{i,j,\mu,\nu}d_{\mu^{\prime}\mu}(r_i)d_{\nu^{\prime}\nu}(r_j)\langle \delta m^{\mu}_i(t) \delta m^{\nu}_j(0)\rangle.
\label{dipole}
\end{equation}
The geometrical factor ${\cal A}^{\mu\nu}({\mib q})$ can be calculated as the Fourier transform of $d_{\mu^{\prime}\mu}(r_i)d_{\nu^{\prime}\nu}(r_j)$ with respect to the wave vector ${\mib q}$.
The nuclear Zeeman axis ($Z$) coincides with the $z$-axis for $H {\mbox {\scriptsize $/\!\!/$}} c$, therefore, $\delta h^{\pm}$ in eq.(\ref{deltah}) is defined as $\delta h^{\pm}= \delta h^x \pm i \delta h^y$. 

 In the case of $H {\mbox{\footnotesize $/\!\!/$}} c$, the expression of ${\cal A}^{\mbox{\tiny $\perp$}}({\mib q})$ is given as
\begin{eqnarray}
& &{\cal A}^{\mbox{\tiny $\perp$}}({\mib q})=\{(\frac{\textstyle 1}{\textstyle r_{\mbox{\tiny $0$}}^3}-\frac{\textstyle a^2}{\textstyle 2r_{\mbox{\tiny $0$}}^5})^2+\frac{\textstyle a^4}{\textstyle 4r_{\mbox{\tiny $0$}}^{10}}\}\cdot\{6+6\cos(2\pi q_c)\} \nonumber \\
&+&\{(\frac{\textstyle 1}{\textstyle r_{\mbox{\tiny $0$}}^3}-\frac{\textstyle a^2}{\textstyle 2r_{\mbox{\tiny $0$}}^5})^2-\frac{\textstyle a^4}{\textstyle 8r_{\mbox{\tiny $0$}}^{10}}\}\cdot\{4\cos(2\pi q_a+2\pi q_b) \nonumber\\
&+&2\cos(2\pi q_a+2\pi q_b-2\pi q_c)+2\cos(2\pi q_a+2\pi q_b+2\pi q_c)  \nonumber \\
&+&4\cos(2\pi q_a)+2\cos(2\pi q_a-2\pi q_c)+2\cos(2\pi q_a+2\pi q_c)  \nonumber \\
&+&4\cos(2\pi q_b)\nonumber+2\cos(2\pi q_b-2\pi q_c)+2\cos(2\pi q_a+2\pi q_c)\}, \nonumber \\
& &
\label{B-Aperpq}
\end{eqnarray}
 where ${\mib q}=(q_a,q_b,q_c)$; $a$, $c$ are the lattice parameters shown in Fig.\ref{Trigonal} and $r_{\mbox{\tiny $0$}}=(a^2/3+c^2/4)^{1/2}$. ${\cal A}^{\mbox{\tiny $/\!\!/$}}({\mib q})$ is given as  
\begin{eqnarray}
& &{\cal A}^{\mbox{\tiny $/\!\!/$}}({\mib q})=\frac{\textstyle 3a^2c^2}{\textstyle 16r_{\mbox{\tiny $0$}}^{10}}\cdot\{6-6\cos(2\pi q_c)-2\cos(2\pi q_a+2\pi q_b) \nonumber\\
&+&\cos(2\pi q_a+2\pi q_b-2\pi q_c)+\cos(2\pi q_a+2\pi q_b+2\pi q_c)  \nonumber \\
&-&2\cos(2\pi q_a)+\cos(2\pi q_a-2\pi q_c)+\cos(2\pi q_a+2\pi q_c)  \nonumber \\
&-&2\cos(2\pi q_b)+\cos(2\pi q_b-2\pi q_c)+\cos(2\pi q_a+2\pi q_c)\}. \nonumber \\
& &
\label{B-Aparaq}
\end{eqnarray}

 In the case of $H \bot c$, the nuclear Zeeman axis ($Z$) coincides with the $x$-axis, therefore, $\delta h^{\pm}$ in eq.(\ref{deltah}) is defined as $\delta h^{\pm}= \delta h^y \pm i \delta h^z$. Then the expressions of ${\cal A}^{xx}({\mib q})$, ${\cal A}^{yy}({\mib q})$ and ${\cal A}^{zz}({\mib q})$ are given as 
\begin{eqnarray}
& &{\cal A}^{xx}({\mib q})=(\frac{\textstyle 3a^4}{\textstyle 16r_{\mbox{\tiny $0$}}^{10}}+\frac{\textstyle 9a^2c^2}{\textstyle 64r_{\mbox{\tiny $0$}}^{10}})\cdot \{
4-4\cos(2\pi q_a+2\pi q_b)\} \nonumber\\
&+&(\frac{\textstyle 3a^4}{\textstyle 16r_{\mbox{\tiny $0$}}^{10}}-\frac{\textstyle 9a^2c^2}{\textstyle 64r_{\mbox{\tiny $0$}}^{10}})\cdot\{4\cos(2\pi q_c)-2\cos(2\pi q_a+2\pi q_b\nonumber\\
&+&2\pi q_c)-2\cos(2\pi q_a+2\pi q_b-2\pi q_c)\},
\label{B-Axxq}
\end{eqnarray}
\begin{eqnarray}
& &{\cal A}^{yy}({\mib q})=\{(\frac{\textstyle 1}{\textstyle r_{\mbox{\tiny $0$}}^3}-\frac{\textstyle a^2}{\textstyle 4r_{\mbox{\tiny $0$}}^5})^2-\frac{\textstyle 3a^2c^2}{\textstyle 64r_{\mbox{\tiny $0$}}^{10}}\}\cdot \{ 4\cos(2\pi q_c) \nonumber\\
&+&2\cos(2\pi q_a+2\pi q_b-2\pi q_c)+2\cos(2\pi q_a+2\pi q_b+2\pi q_c)\}\nonumber\\
&+&\{(\frac{\textstyle 1}{\textstyle r_{\mbox{\tiny $0$}}^3}-\frac{\textstyle a^2}{\textstyle 4r_{\mbox{\tiny $0$}}^5})^2+ \frac{\textstyle 3a^2c^2}{\textstyle 64r_{\mbox{\tiny $0$}}^{10}}\} \cdot \{4+4\cos(2\pi q_a+2\pi q_b)\}\nonumber\\
&+&\{(\frac{\textstyle 1}{\textstyle r_{\mbox{\tiny $0$}}^3}-\frac{\textstyle a^2}{\textstyle r_{\mbox{\tiny $0$}}^5})^2+\frac{\textstyle a^2c^2}{\textstyle 32r_{\mbox{\tiny $0$}}^{10}}\} \cdot 2+\{(\frac{\textstyle 1}{\textstyle r_{\mbox{\tiny $0$}}^3}-\frac{\textstyle a^2}{\textstyle r_{\mbox{\tiny $0$}}^5})^2-\frac{\textstyle a^2c^2}{\textstyle 32r_{\mbox{\tiny $0$}}^{10}}\}\nonumber\\ 
&\cdot& 2\cos(2\pi q_c)+\{(\frac{\textstyle 1}{\textstyle r_{\mbox{\tiny $0$}}^3}-\frac{\textstyle a^2}{\textstyle 4r_{\mbox{\tiny $0$}}^5})\cdot(\frac{\textstyle 1}{\textstyle r_{\mbox{\tiny $0$}}^3}-\frac{\textstyle a^2}{\textstyle r_{\mbox{\tiny $0$}}^5})-\frac{\textstyle 3a^2c^2}{\textstyle 32r_{\mbox{\tiny $0$}}^{10}}\}\nonumber\\
&\cdot&\{4\cos(2\pi q_a)+4\cos(2\pi q_b)\} +\{(\frac{\textstyle 1}{\textstyle r_{\mbox{\tiny $0$}}^3}-\frac{\textstyle a^2}{\textstyle 4r_{\mbox{\tiny $0$}}^5})\cdot(\frac{\textstyle 1}{\textstyle r_{\mbox{\tiny $0$}}^3}-\frac{\textstyle a^2}{\textstyle r_{\mbox{\tiny $0$}}^5})\nonumber\\
&+&\frac{\textstyle 3a^2c^2}{\textstyle 32r_{\mbox{\tiny $0$}}^{10}}\}\cdot\{2\cos(2\pi q_a-2\pi q_c)+2\cos(2\pi q_a+2\pi q_c)\nonumber\\ 
&+&2\cos(2\pi q_b-2\pi q_c)+2\cos(2\pi q_b+2\pi q_c)\},
\label{B-Ayyq}
\end{eqnarray}
\begin{eqnarray}
& &{\cal A}^{zz}({\mib q})=\{(\frac{\textstyle 1}{\textstyle r_{\mbox{\tiny $0$}}^3}-\frac{\textstyle 3c^2}{\textstyle 16r_{\mbox{\tiny $0$}}^5})^2-\frac{\textstyle 3a^2c^2}{\textstyle 64r_{\mbox{\tiny $0$}}^{10}}\}\cdot \{4\cos(2\pi q_c) \nonumber\\
&+&2\cos(2\pi q_a+2\pi q_b-2\pi q_c)+2\cos(2\pi q_a+2\pi q_b+2\pi q_c)\}\nonumber\\
&+&\{(\frac{\textstyle 1}{\textstyle r_{\mbox{\tiny $0$}}^3}-\frac{\textstyle 3c^2}{\textstyle 16r_{\mbox{\tiny $0$}}^5})^2+\frac{\textstyle 3a^2c^2}{\textstyle 64r^{10}_{\mbox{\tiny $0$}}}\} \cdot \{4+4\cos(2\pi q_a+2\pi q_b)\}\nonumber\\
&+&\{(\frac{\textstyle 1}{\textstyle r_{\mbox{\tiny $0$}}^3}-\frac{\textstyle 3c^2}{\textstyle 16r_{\mbox{\tiny $0$}}^5})^2+\frac{\textstyle 3a^2c^2}{\textstyle 16r_{\mbox{\tiny $0$}}^{10}}\}\cdot2+\{(\frac{\textstyle 1}{\textstyle r_{\mbox{\tiny $0$}}^3}-\frac{\textstyle 3c^2}{\textstyle 16r_{\mbox{\tiny $0$}}^5})^2-\frac{\textstyle 3a^2c^2}{\textstyle 16r_{\mbox{\tiny $0$}}^{10}}\}\nonumber\\
&\cdot&2\cos(2\pi q_c)+\{(\frac{\textstyle 1}{\textstyle r_{\mbox{\tiny $0$}}^3}-\frac{\textstyle 3c^2}{\textstyle 16r_{\mbox{\tiny $0$}}^5})^2-\frac{\textstyle 3a^2c^2}{\textstyle 32r_{\mbox{\tiny $0$}}^{10}}\}\cdot\{4\cos(2\pi q_a) \nonumber\\
&+&4\cos(2\pi q_b)\}+\{(\frac{\textstyle 1}{\textstyle r_{\mbox{\tiny $0$}}^3}-\frac{\textstyle 3c^2}{\textstyle 16r_{\mbox{\tiny $0$}}^5})^2+\frac{\textstyle 3a^2c^2}{\textstyle 32r_{\mbox{\tiny $0$}}^{10}}\} \nonumber\\ 
&\cdot&\{2\cos(2\pi q_a-2\pi q_c)+2\cos(2\pi q_a+2\pi q_c) \nonumber\\ 
&+&2\cos(2\pi q_b-2\pi q_c)+2\cos(2\pi q_b+2\pi q_c)\}.
\label{B-Azzq}
\end{eqnarray}

\end{document}